\documentclass{aa}
\usepackage[varg]{txfonts}

\usepackage{array, multirow, graphicx,natbib,txfonts,color,tabularx,amsmath}
\usepackage{lscape,float,placeins}
\usepackage{threeparttable}
\bibpunct{(}{)}{;}{a}{}{,}

\newcommand{\ang}{$\rm \AA$}
\newcommand{\FeH}{$\left[\mbox{Fe/H}\right]$}
\newcommand{\kms}{km\,s$^{-1}$}

\begin{document}

\title{Atomic diffusion and mixing in old stars IV: Weak abundance trends in the globular cluster NGC\,6752\thanks{Based on data collected at the ESO telescopes under programmes 079.D-0645\, (A) and 81.D-0253\,(A)}}

\author{Pieter Gruyters \inst{1} \and Andreas J. Korn \inst{1} \and Olivier Richard \inst{3} \and Frank Grundahl \inst{2} \and Remo Collet \inst{5,6} \and Lyudmila I. Mashonkina \inst{4} \and Yeisson Osorio \inst{1} \and Paul S. Barklem \inst{1}}

\offprints{pieter.gruyters@physics.uu.se}

\institute{Department of Physics and Astronomy, Division of Astronomy and Space Physics, Uppsala University, Box 516, 75120 Uppsala, Sweden \and Stellar Astrophysics Centre, Department of Physics and Astronomy, \AA rhus University, Ny Munkegade 120, DK-8000 \AA rhus C, Denmark. \and LUPM, Universit\'e Montpellier II, CNRS, CC072, Place E. Bataillon, 34095 Montpellier Cedex, France \and Institute of Astronomy, Russian Academy of Science, 119017 Moscow, Russia \and Research School of Astronomy \& Astrophysics, Cotter Road, Weston ACT 2611, Australia \and Centre for Star and Planet Formation, Natural History Museum of Denmark / Niels Bohr Institute,
University of Copenhagen, {\O}ster Voldgade 5-7, DK--1350 Copenhagen, Denmark }

\date{Received  / Accepted}

\authorrunning{P. Gruyters}
\titlerunning{Atomic Diffusion in NGC6752}

\abstract
{Atomic diffusion in stars can create systematic trends of surface abundances with evolutionary stage. Globular clusters offer useful laboratories to put observational constraints on this theory as one needs to compare abundances in unevolved and evolved stars, all drawn from the same stellar population. }
{Atomic diffusion and additional mixing has been shown to be at work in the globular cluster NGC\,6397 at a metallicity of \FeH $\sim-2.1$. We investigate possible abundance trends in Li, Mg, Ca, Ti, Sc, and Fe with evolutionary stage in another globular cluster NGC\,6752 at a metallicity of \FeH $\sim-1.6$. This in order to better constrain stellar structure models including atomic diffusion and additional mixing. }
{We performed a differential abundance analysis on VLT/FLAMES-UVES data of 16 stars in four groups between the turnoff point and the red giant branch. Continuum normalisation of the stellar spectra was performed in an automated way using DAOSPEC. Differential abundances relative to the sun were derived by fitting synthetic spectra to individual lines in the stellar spectrum.}
{ We find weak systematic abundance trends with evolutionary phase for Fe, Sc, Ti, and Ca. The individual trends are weaker than the trends in NGC\,6397 and only significant at the 1-$\sigma$ level. However, the combined trend shows a significance on the 2-$\sigma$ level. The trends are best explained by stellar-structure models including atomic diffusion with more efficient additional mixing than needed in NGC\,6397. The model allows to correct for sub-primordial stellar lithium abundances of the stars on the Spite plateau.}
{Abundance trends for groups of elements, differently affected by atomic diffusion and additional mixing, are identified. Although the significance of the trends is weak, they all seem to indicate that atomic diffusion is operational along the evolutionary sequence of NGC\,6752. The trends are weaker than those observed in NGC\,6397, which is perhaps due to more efficient mixing. Using models of atomic diffusion including efficient additional mixing, we find a diffusion-corrected primordial lithium abundance of $\log \varepsilon$(Li) = $2.58\pm0.10$, in agreement with WMAP-calibrated Big-Bang nucleosynthesis predictions within the mutual 1-$\sigma$ uncertainties.}


\maketitle


\section{Introduction}\label{sect:intro}
Globular clusters (GCs) are considered to be the oldest stellar aggregates in our Galaxy. They constitute, in many respects, homogeneous stellar populations at least in terms of distance, age, and iron-peak elemental abundances. Obtaining chemical abundances of individual stars in a GC therefore gives important clues to the physical processes at work in these stars. One example of this can be found in the paper by \citet{Korn2007}. By analysing the chemical abundances of stars in different evolutionary stages in the metal-poor (\FeH \footnote{We adopt here the usual spectroscopic notations that [X/Y] $\equiv$ log\,(N$_X$/N$_Y$)$_{*}$--log\,(N$_X$/N$_Y$)$_{\odot}$, and that log\,$\varepsilon$(X)$\equiv$ log\,(N$_X$/N$_H$)+12 for elements X and Y. We assume also that metallicity is equivalent to the stellar [Fe/H] value.}\,$\sim-2.1$) GC NGC\,6397 they were able to show the existence of systematic differences in the surface abundances of these stars. Using FLAMES-UVES, Korn et al. collected spectra for four groups of stars located between the main sequence (MS) turnoff point (TOP) and the red giant branch (RGB). The analysis revealed significant trends in surface abundances with effective temperature $(T_{\mbox{\scriptsize eff}})$ and surface gravity (log\,$g$) in the atmospheres of TOP stars compared to RGB stars. Atomic diffusion with some degree of additional mixing (hereafter referred to as AddMix) was invoked to explain these differences. \\
Atomic diffusion (AD) is a continuous process which modifies the chemical composition of the surface layers during the MS lifetime of a star. As the star evolves to the RGB the effect of AD disappears as the star's deep outer convection zone restores the original composition in the atmosphere (with the notable exception of lithium). However, stellar models including only AD result in too strong trends of metals between the surface layers of TOP and RGB stars. To counteract uninhibited AD, an ad-hoc parameterisation of AddMix is incorporated in these models. Mixing hinders the downward diffusion of elements. Incorporating more efficient mixing will then produce flattened abundance trends in stellar-structure models with AD.

Because stars within a GC are expected to have the same age and original chemical composition in terms of iron-peak elements, they offer suitable test cases for AD studies. Although the inclusion of AddMix in a layer just below the convective envelope is remarkably successful in describing the observed abundance trends in NGC\,6397, the description applied is in no way unique and physically satisfying. \\
To get a better understanding of the involved physics and put additional constraints on the dependence of AddMix on stellar parameters such as metallicity, we here present the results of a study similar to the one presented in \citet{Korn2007} on another metal-poor GC, NGC\,6752 at a metallicity \FeH\, $\sim -1.6$.\\

Throughout the last decennia, NGC\,6752 has been studied in detail by quite a number of authors. In 2001, \citet{Gratton2001} published a study in which they derived chemical abundances for 9 TOP and 9 bRGB stars in NGC\,6752 to study O-Na and Mg-Al anticorrelations. They also deduced Fe abundances for the stars and found no indication of variations between TOP and bRGB stars. \citet{James2004} rederived Fe abundances for the 9 TOP and 9 base-RGB (hereafter bRGB) stars in the Gratton sample while performing an abundance analysis for heavy elements in the stars. They again found no Fe abundance difference between the two groups. Over the recent years, the Padova group has published a series of papers (i.e. \citealt{Carretta2006,Carretta2007b,Carretta2007a,Carretta2007c}; \citealt{Gratton2006,Gratton2007}) in which they analyse anticorrelations in a sample of 19 GCs including NGC\,6752 and NGC\,6397. From the study they conclude that GCs are characterised by a Na-O anticorrelation. They contain a first and second generation of stars where the first generation constitutes roughly a third of the stars in a GC. By studying the anticorrelations in detail they come to the conclusion that GCs must have been built up in at least two star-forming phases \citep{Carretta2010}. A recent paper by \citet{Carretta2012} presented evidence for three distinct stellar populations in NGC\,6752. Being only interested in anticorrelations, none of their papers addresses possible variations in Fe abundance with evolutionary phase. In this paper we revisit NGC\,6752 from a diffusion point of view. Preliminary results were already presented in \citet{Korn2010} indicating a small but systematic abundance difference in iron between TOP and RGB stars: $\Delta$ log\,$\varepsilon$(Fe) = $-0.10 \pm 0.03$ between TOP and RGB stars.\\
The paper is organised as follows: In Sect.\,2 the observations are discussed. The data reduction techniques and methodology are given in Sect.\,3 while in Sect.\,4 we present the photometric stellar parameters and spectroscopic abundance analysis. In Sect.\,5, abundance trends are compared with predictions from stellar evolution models including AD and AddMix. Finally, in Sect.\,6 we present our conclusions.



\section{Observations and data reduction}\label{sect:obs}

The observations were obtained during ESO period 79 (program ID: 079.D-0645 (A)) and period 81 (programme ID: 081.D-0253 (A)) using the multi-object spectrograph FLAMES-UVES mounted on the ESO VLT-UT2 Kueyen \citep{Pasquini2003}. During a total exposure time of about 70\,hrs we obtained spectra for four groups of stars in different evolutionary stages on and between the TOP and the RGB with single exposure times not exceeding 4215\,s. In total we have data for 16 stars, five TOP stars, one subgiant branch star (SGB), four bRGB and six RGB stars, covering a range in $T_{\mbox{{\scriptsize eff}}}$ between $5000$\,K and $6100$\,K and log\,$g$ between 2.5 and 4. The wavelength coverage of the spectra is 4800--6800\,\ang\ at a spectral resolution of $R=47,000$. The spectra were reduced using the FLAMES-UVES data reduction pipeline. An overview of the data can be found in Table\,\ref{Tab:log}.\\

\begin{table}
\caption{Observational data of the NGC\,6752 stars observed with FLAMES-UVES.}
\label{Tab:log}      
\centering          
\begin{tabular}{lccccc} 
\hline\hline
Group & Star ID & $V$ & $n_{\mbox{{\scriptsize exp}}}$  &$t_{\mbox{\scriptsize exp}}$ & S/N$_{\mbox{\scriptsize tot}}$ \\
 & & & & (hr) & (pixel$^{-1}$) \\ 
\hline                    
 TOP &  3988 & 17.053 & 26 & 30.4 & 37 \\  
 & 4096 & 17.089 & 26 & 30.4 & 36 \\
 & 4138 & 17.084 & 26 & 30.4 & 23 \\
 & 4383 & 17.113 & 26 & 30.4 & 34 \\
 & 4428 & 17.142 & 26 & 30.4 & 35 \\
 SGB & 3081 & 16.809 & 34 & 39.8 & 51 \\
 bRGB & 1406 & 15.903 & 13 & 16.9 & 63 \\
 & 1483 & 15.957 & 13 & 16.9 & 59 \\
 & 1522 & 15.987 & 13 & 16.9 & 65 \\
 & 1665 & 16.044 & 13 & 16.9 & 56 \\
 RGB & 540 & 14.451 & 2 & 1.7 & 60 \\
 & 548 & 14.491 & 2 & 1.7 & 52 \\
 & 566 & 14.516 & 2 & 1.7 & 63 \\
 & 574 & 14.545 & 2 & 1.7 & 65 \\
 & 581 & 14.568 & 2 & 1.7 & 62 \\
 & 588 & 14.570 & 2 & 1.7 & 64 \\
\hline
\end{tabular}
\tablefoot{S/N$_{\mbox{\scriptsize tot}}$ refers to the signal-to-noise ratio per pixel in the co-addition of the $n_{\mbox{{\scriptsize exp}}}$ spectra after rebinning by a factor of two (retaining the full resolving power of FLAMES-UVES).}
\end{table}

After the reduction process, continuum normalisation was performed. This was done using DAOSPEC \citep{Stetson2008}, a publicly available tool which calculates a fit to the continuum of a spectrum and measures the equivalent width (EW) of lines. Since DAOSPEC is an automated tool it assures full reproducibility of the continuum normalisation, removing potential biases introduced by manual placement of the continuum. To verify DAOSPEC produces reliable results, we performed some tests similar to the ones presented in \citet{Stetson2008} on observed data. Most of their results can be confirmed. However, there is one systematic effect that we seem to encounter for our dataset. The continuum seems to be placed too high in low signal-to-noise (S/N) ($<50$) spectra. This finding together with the result that the difference in EW between the true value and the value measured from a spectrum where the continuum is misplaced by a given amount, is approximately constant  \citep{Stetson2008}, led us to co-add the five TOP spectra into a single spectrum with S/N of about 60. That way we base our abundance analysis on a homogeneous set of spectra in terms of S/N and minimise the uncertainty due to continuum placement.\\

\begin{figure}
\begin{center}
\includegraphics[width=1\columnwidth]{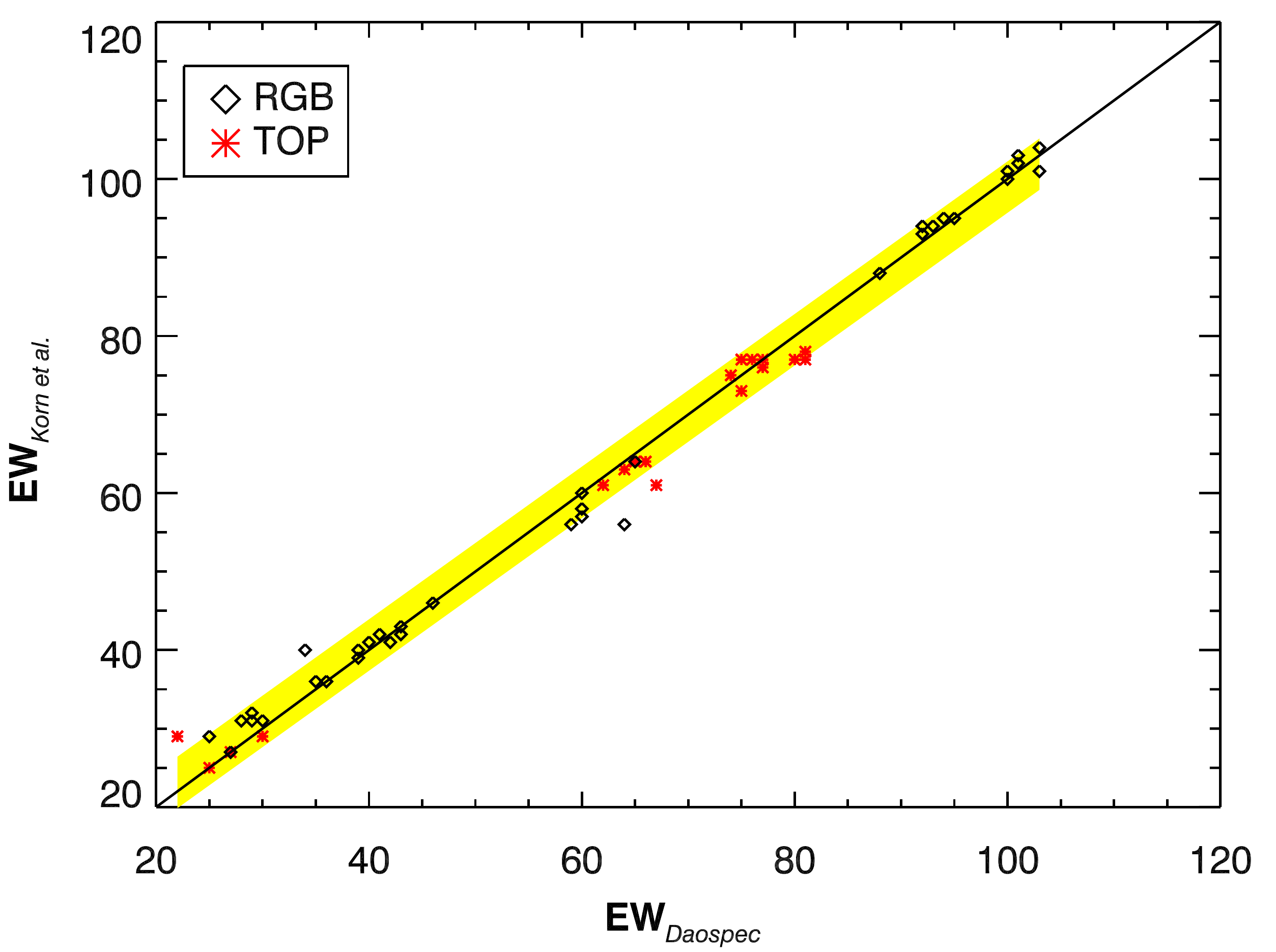}
\caption{Comparison between the EWs of Fe\,{\scriptsize II} lines in TOP and RGB stars in NGC\,6397 measured by DAOSPEC and \citet{Korn2007}. The yellow bar around the solid line gives the 1-$\sigma$ zone around the one-to-one correspondence.}\label{Fig:EW_comp}
\end{center}
\end{figure}

We checked the EWs as measured by DAOSPEC on UVES spectra by comparing the EWs of the dataset of \citet{Korn2007} with the EWs derived by DAOSPEC for this same dataset. The result is shown in Fig.\,\ref{Fig:EW_comp} where we plot the EW of Fe\,{\scriptsize II} lines measured from the same UVES spectra of TOP and RGB stars in the GC NGC\,6397. The agreement is excellent: on average the difference (DAOSPEC--Korn et al.) is $-0.09\pm0.04$ m\AA\ with $\sigma=2.50$\,m\AA\ from 70 lines. A linear regression between the two sets of measurements gives EW$_{\mbox{\scriptsize DAOSPEC}}=0.97\times$ EW$_{\mbox{\scriptsize Korn et al.}}+1.75$ with rms=2.39\,m\AA\ and a correlation coefficient $r$=0.99. We thus confirm the abundance trend for NGC\,6397 with the tools presented here:
\begin{center}
$\Delta$ log\,$\varepsilon$(Fe)$_{\mbox{\scriptsize NGC6397}}=0.15\pm0.05$
\end{center}
for a $\Delta T$ of 1108 $\pm$ 27\,K and $\Delta$ log $g$ of $1.38 \pm 0.06$ between TOP and RGB stars on the $(b-y)$ effective temperature scale by \citet{Alonso1996,Alonso1999}.

\section{Analysis}
\subsection{Photometry}

We use similar photometric data to that used by \citet{Korn2006, Korn2007}. It consists of $uvby$ Str\"omgren photometry and we employ effective temperature estimates on the $(b-y)$ colour. The data reduction and photometric calibration procedures have been discussed in the papers on GC $uvby$ photometry by \citet{Grundahl1998,Grundahl1999}. We have estimated the photometric zero point errors to be 0\fm01 in both $b$ and $y$.

\subsection{Chemical abundances}
AD is expected to affect all chemical elements. However, not all elements are suited to study the effect by comparing groups of stars in different evolutionary phases. Helium is not observable in solar-type stars. Lithium is heavily affected by internal cluster pollution in NGC\,6752 \citep{Pasquini2005,Shen2010}. Beryllium and boron require near-UV spectroscopy and are significantly processed in RGB stars making them unusable in a comparative study like ours. Carbon and nitrogen are dredge-up elements and thus the surface abundances may reflect contamination of the atmosphere by the processed material from the core rather than the original compositions. Other elements such as oxygen, sodium, magnesium and aluminium suffer from anticorrelations \citep{Kraft1997} as a result of external pollution \citep{Gratton2001}. These restrictions limit us to elements such as silicon, calcium, scandium, titanium and iron-group elements for differential analyses between groups of stars.  
 
The effective temperatures ($T_{\mbox{{\scriptsize eff}}}$), used for deriving the chemical surface abundances, are deduced through $(b-y)$ photometric data using E$(b-y) = 0.03$ and $(m-M)_V=13.30$. The photometric calibration was kindly provided by L. Casagrande (private comm. 2009) and is valid for dwarfs and giants. Using photometry in conjunction with the fact that all stars have the same distance modulus, the surface gravity can be derived. The stellar masses needed were obtained by interpolation in a 13.5Gyr  isochrone from \citet*{Richard2005}. The masses varied from 0.810 to 0.835$M_{\odot}$.  An uncertainty of $\pm$100\,K  on $T_{\mbox{{\scriptsize eff}}}$ leads to an uncertainty of 0.07\,dex in abundance for neutral species, a fairly large error given the small abundance differences we are looking for \citep{Korn2010}. However, by analysing lines of ionised species, which are predominately gravity sensitive, we can circumvent this problem since a shift of $\pm$100\,K in $T_{\mbox{{\scriptsize eff}}}$ leads to no appreciable shift in log\,$g$ and thus we may assume negligible influence on the abundances. This strategy limits us to the elements iron, titanium and scandium. 
Stellar structure models by \citet*{Richard2005} including AD with AddMix parametrised by T6.0, predict the strongest variation for silicon and magnesium [$\Delta\mbox{log}\,\varepsilon(X)=\mbox{log}\,\varepsilon(X)_{\mbox{\scriptsize RGB}}-\mbox{log}\,\varepsilon(X)_{\mbox{\scriptsize TOP}}$ = 0.2]. Caution should be exercised when looking at these elements as it has been shown that there exists a correlation between silicon and aluminium which is consistent with the abundance anomalies being synthesized via proton captures at high temperatures \citep{Young2005}. As no silicon lines were detected in our spectra, we analyse magnesium, but caution that magnesium correlates with oxygen \citep{Gratton2001}.\\

With the stellar parameters deduced from photometry, it becomes straightforward to deduce abundances and potential abundance differences. Since we are interested in abundance differences, we perform a line-by-line differential analysis where we compare the abundances derived from a stellar spectrum relative to the Sun. In that way we minimise any potential biases introduced by choosing specific atomic data. An exception are the Sc\,{\scriptsize II} lines, which are affected by hyperfine structure (HFS) that was not taken into account in our calculations. To our knowledge, no HFS data is available for the investigated Sc\,{\scriptsize II} lines at 5031, 5526, and 5657\,\AA. If wavelength separation between the HFS components of these lines is as large as that for Sc\,{\scriptsize II} 4415\,\AA\ \citep*{McWilliam1995}, ignoring HFS effects leads to a slight (0.03-0.04\,dex) overestimation in Sc abundance for the RGB stars, while it has nearly no influence on the Sc abundance of the TOP stars. \\

 Abundances are derived using SIU \citep{Reetz1991}, a visualisation tool to compare observed and theoretical spectra. SIU is equipped with a built-in line-synthesis module. The module uses one-dimensional hydrostatic model atmospheres in local thermodynamical equilibrium (LTE) with an opacity distribution function (ODF) representation of line opacity (MAFAGS, \citealp{Fuhrmann1997,Grupp2004}). In our differential analysis, we use the high-S/N Kitt-Peak Solar Atlas \citep{Kurucz1984} as solar spectrum to compare our line abundances to. As a result of the highly differential character of the star-to-star analysis, the results are almost completely independent of the model atmospheres used, and using more detailed MARCS models \citep{Gustafsson2008} would yield the same results. However, although we are working with differential indicators, there can be some effects due to departures from LTE as excitation, ionisation and collisions are inherently different in dwarfs and giant stars \citep{Korn2007}. By using lines arising from transitions in Sc\,{\scriptsize II}, Ti\,{\scriptsize II} and Fe\,{\scriptsize II} we partially circumvent the NLTE problem for these elements since these lines are believed to be formed under near--LTE conditions and constitute the dominant ionisation stages of the respective element. Mg abundances are derived from the Mg\,{\scriptsize I} lines at 5528 and 5711\,\AA. As these lines of neutral magnesium can be affected by departures from LTE, we apply NLTE corrections to abundances derived from LTE line formation. Like in NGC\,6397 we include two more elements in our analysis: Calcium as inferred from the Ca\,{\scriptsize I}\,$\lambda\lambda$6169.5, 6493.7 lines and barium for which we use the Ba\,{\scriptsize II} line at 6141.7\,\AA. The Ba\,{\scriptsize II} $\lambda$6496.9 line is contaminated by telluric absorption. We therefore disregard this line from our analysis. As for Mg we apply NLTE corrections to the Ca and Ba abundances.\\

\newpage
\begin{landscape}
\begin{table}
\caption{Stellar parameters using b-y photometry and derived abundances for stars in NGC6752.}\label{Tab:Full-Results}
\centering
\begin{tabular}{clccccccccc}\hline\hline
Group & Star ID & $T_{\mbox{\scriptsize eff}}$ & log\,$g$ & $\xi$ & [Fe/H]$\pm1\sigma$ & log\,$\varepsilon$(Mg) & log\,$\varepsilon$(Sc)  & log\,$\varepsilon$(Ti) & log\,$\varepsilon$(Ca) & log\,$\varepsilon$(Ba) \\ 
 & & (K) & (cgs) & (km\,s$^{-1}$) & (dex) & LTE & LTE & LTE & LTE & LTE \\ \hline
RGB & 540 & 5000 & 2.54 & 1.50 & $-1.65\pm0.12$ & $6.21\pm0.07$ & $1.29\pm0.10$ & $3.43\pm0.05$ & $4.90\pm0.04$ & $0.77\pm0.03$\\
 	& 548 & 5082 & 2.59 & 1.50 & $-1.65\pm0.12$ & $6.02\pm0.07$ & $1.37\pm0.09$ & $3.47\pm0.06$ & $4.98\pm0.04$ & $0.73\pm0.03$\\
 	& 566 & 5000 & 2.56 & 1.40 & $-1.68\pm0.02$ & $6.22\pm0.07$ & $1.34\pm0.08$ & $3.48\pm0.05$ & $5.00\pm0.04$ & $0.65\pm0.03$\\
 	& 574 & 5045 & 2.60 & 1.50 & $-1.64\pm0.08$ & $6.25\pm0.07$ & $1.32\pm0.08$ & $3.44\pm0.06$ & $4.96\pm0.04$ & $0.65\pm0.03$\\
 	& 581 & 5040 & 2.60 & 1.40 & $-1.67\pm0.11$ & $6.24\pm0.07$ & $1.31\pm0.09$ & $3.47\pm0.07$ & $5.01\pm0.04$ & $0.65\pm0.03$\\
 	& 588 & 5018 & 2.59 & 1.50 & $-1.66\pm0.05$ & $6.16\pm0.07$ & $1.33\pm0.10$ & $3.43\pm0.05$ & $4.99\pm0.04$ & $0.75\pm0.03$\\ \hline
RGB$_{\mbox{mean}}$ & ... & $5031\pm32$ & $2.58\pm0.02$ & $1.45$ & $-1.66\pm0.04$ & $6.22\pm0.04$ & $1.33\pm0.03$ & $3.45\pm0.02$ & $4.97\pm0.04$ & $0.70\pm0.06$\\ \hline
bRGB & 1406 & 5234 & 3.22 & 1.40 & $-1.62\pm0.05$ & $6.24\pm0.07$ & $1.35\pm0.07$ & $3.46\pm0.06$ & $4.99\pm0.04$ & $0.73\pm0.03$ \\
	  & 1483 & 5253 & 3.25 & 1.40 & $-1.74\pm0.04$ & $6.18\pm0.07$ & $1.32\pm0.10$ & $3.40\pm0.09$ & $4.90\pm0.04$ & $0.72\pm0.03$ \\
	  & 1522 & 5288 & 3.28 & 1.40 & $-1.67\pm0.05$ & $6.30\pm0.07$ & $1.31\pm0.09$ & $3.45\pm0.06$ & $4.97\pm0.04$ & $0.73\pm0.03$ \\
	  & 1665 & 5248 & 3.28 & 1.40 & $-1.64\pm0.07$ & $6.28\pm0.07$ & $1.28\pm0.10$ & $3.50\pm0.05$ & $4.87\pm0.05$ & $0.72\pm0.03$ \\ \hline
bRGB$_{\mbox{mean}}$ & ... & $5256\pm23$ & $3.26\pm0.03$ & $1.40$ & $-1.67\pm0.03$ & $6.25\pm0.05$ & $1.32\pm0.03$ & $3.45\pm0.03$ & $4.93\pm0.06$ & $0.72\pm0.01$ \\ \hline
SGB & 3081 & 5793 & 3.79 & 1.50 & $-1.68\pm0.10$ & $6.07\pm0.10$ & $1.21\pm0.09$ & $3.28\pm0.08$ & $4.76\pm0.07$ & $0.48\pm0.06$ \\ \hline
TOP & 3988 & 6089 & 3.99 & 1.95 & $-1.79\pm0.12$ & $6.06\pm0.11$ &	        &       & $5.00\pm0.10$ & $0.91\pm0.04$ \\
 	& 4096 & 6196 & 4.03 & 1.95 & $-1.73\pm0.07$ & $6.03\pm0.11$ &		&	& $5.00\pm0.10$ & $0.45\pm0.07$ \\
 	& 4138 & 6049 & 3.98 & 1.95 & $-1.67\pm0.02$ & $6.06\pm0.11$ & no reliable	& abundances & $4.99\pm0.10$ & $0.43\pm0.07$ \\
   	& 4383 & 6032 & 3.99 & 1.80 & $-1.68\pm0.07$ & $6.13\pm0.11$ &		&	& $4.94\pm0.10$ & $0.43\pm0.07$ \\
    	& 4428 & 6001 & 3.99 & 1.90 & $-1.75\pm0.08$ & $6.10\pm0.11$ &		&	& $4.76\pm0.10$ & $0.36\pm0.07$ \\ \hline
TOP$_{\mbox{mean}}$ & ... & $6068\pm57$ & $4.00\pm0.01$ & $2.00$  & $-1.72\pm0.07$ & $6.08\pm0.04$ &  &  & $4.93\pm0.10$ & $0.42\pm0.04$ \\ \hline
$\Delta$[TOP-RGB] & ... & $1037\pm65$ & $1.42\pm0.02$ & & $0.06\pm0.08$ & $0.14\pm0.06$ &  &  & $0.04\pm0.11$ & $0.28\pm0.07$ \\
$\Delta$[TOP-bRGB] & ... & $812\pm61$ & $0.74\pm0.03$ & & $0.05\pm0.08$ & $0.17\pm0.06$ &  &  & $0.00\pm0.12$ & $0.30\pm0.04$ \\
\hline\hline
\end{tabular}
\tablefoot{All abundances are derived fully differential to the Sun. Uncertainties on stellar parameters correspond to the standard deviations. Uncertainties on abundances correspond to standard deviations in case of the individual stars, and propagated errors ($\sum\sqrt{\sigma^2}/\sqrt{N(N-1)}$) for the mean values. For Fe we used 5 Fe\,{\scriptsize II} lines: 4923, 5197, 5234, 5316, and 5362\AA, for Mg 2 Mg\,{\scriptsize I} lines: 5528 and 5711\AA, for Sc we used 3 Sc\,{\scriptsize II} lines: 5031, 5526 and 5657\AA, for Ti we used 5 Ti\,{\scriptsize II} lines: 5129, 5154, 5185, 5226 and 5336\AA, for Ca, we used the Ca\,{\scriptsize I} 6169 and 6493\AA\ lines and for Ba, only the Ba\,{\scriptsize II} at 6141\AA\ was used since the Ba\,{\scriptsize II} at 6496\AA\ was contaminated by telluric absorption. The average Mg abundance for the RGB group is derived disregarding star 548 and the average Ba abundance for the TOP group is derived by disregarding the Ba-rich star 3988 (see text for details). Note that the Sc abundances for the bRGB and RGB stars can be slightly overestimated due to ignoring HFS.}
\end{table}
\end{landscape} 

Since we deduce the elemental abundances from lines of different strength, we set the microturbulence so that the Fe\,{\scriptsize II} line abundances show no trend with line strength. The number of Fe\,{\scriptsize II} lines we were able to fit varies with evolutionary state: for the RGB and bRGB stars we fitted nine lines while for the SGB star we could fit eight lines. Only five lines could be measured in the TOP stars. The line wavelengths can be found in Table\,\ref{Tab:lines}. Once the microturbulence is set, we calculate the mean Fe abundance using only the lines which were measured in all groups of stars, i.e. we derive the Fe abundance using five Fe\,{\scriptsize II} lines. 

\begin{table}[ht]
\caption{Iron lines used in the microturbulence determination.}\label{Tab:lines}
\begin{center}
\begin{tabular}{ccccc}\hline\hline
Line & TOP & SGB & bRGB & RGB \\ \hline
Fe\,{\scriptsize II} $\lambda4923.9$ & X & X & X & X \\ 
Fe\,{\scriptsize II} $\lambda5197.5$ & X & X & X & X \\ 
Fe\,{\scriptsize II} $\lambda5234.6$ & X & X & X & X \\ 
Fe\,{\scriptsize II} $\lambda5264.8$ &  & X & X & X \\ 
Fe\,{\scriptsize II} $\lambda5284.1$ &  & X & X & X \\ 
Fe\,{\scriptsize II} $\lambda5316.6$ & X & X & X & X \\ 
Fe\,{\scriptsize II} $\lambda5325.5$ & & X & X & X \\ 
Fe\,{\scriptsize II} $\lambda5362.8$ & X & X & X & X \\ 
Fe\,{\scriptsize II} $\lambda5425.2$ & & & X & X \\ \hline\hline
\end{tabular}
\end{center}
\end{table}

\begin{table*}
\caption{Mean stellar parameters using ($b-y$) photometry and derived mean abundances for stars in NGC\,6752 from the co-added spectra.  }\label{Tab:Mean-Results}
\centering
\begin{tabular}{lccccccccc}\hline\hline
{\sc Group} & $T_{\mbox{\scriptsize eff}}$ & log\,$g$ & $\xi$ & [Fe/H] & log\,$\varepsilon$(Mg)\tablefootmark{a} & log\,$\varepsilon$(Sc)\tablefootmark{b}  & log\,$\varepsilon$(Ti)\tablefootmark{c}  & log\,$\varepsilon$(Ca)\tablefootmark{d}  & log\,$\varepsilon$(Ba)\tablefootmark{e} \\ 
& (K) & (cgs) & (km\,s$^{-1}$) & (dex) & NLTE & LTE & LTE & NLTE & NLTE \\ \hline
TOP$_{\mbox{\scriptsize ave}}$ & $6068$ & $4.00$ & $2.00$  & $-1.74$ & $6.01$ & $1.23$ & $3.39$ & 4.88 & 0.46 \\
SGB$_{\mbox{\scriptsize ave}}$ & $5793$ & $3.79$ & $1.50$ & $-1.68$ & $6.05$ & $1.21$ & $3.28$ & 4.76 & 0.44 \\
bRGB$_{\mbox{\scriptsize ave}}$ & $5256$ & $3.26$ & $1.40$ & $-1.65$ & $6.16$ & $1.30$ & $3.44$ & 4.93 & 0.48 \\
RGB$_{\mbox{\scriptsize ave}}$ & $5031$ & $2.58$ & $1.45$ & $-1.64$ & $6.12$ & $1.32$ & $3.47$ & 4.96 & 0.42\\
\hline\hline
\end{tabular}
\tablefoot{
\tablefoottext{a}{Based on Mg\,{\scriptsize I} $\lambda\lambda$5528 and 5711. RGB$_{\mbox{\scriptsize ave}}$ Mg abundance derived by disregarding star 548 and using $T_{\mbox{\scriptsize eff}}=5021$K, log\,$g$=2.58, and $\xi=1.45$\,\kms.}
\tablefoottext{b}{Based on Sc\,{\scriptsize II} $\lambda\lambda$5031, 5526 and 5657. Note that the Sc abundances for the bRGB and RGB stars can be slightly overestimated due to ignoring HFS.}
\tablefoottext{c}{Based on Ti\,{\scriptsize II} $\lambda\lambda$5129, 5154, 5185, 5226 and 5336.}
\tablefoottext{d}{Based on Ca\,{\scriptsize I} $\lambda\lambda$6169 and 6493.}
\tablefoottext{e}{Based on Ba\,{\scriptsize II} $\lambda\lambda$6141. TOP$_{\mbox{\scriptsize ave}}$ Ba abundance derived by disregarding star 3988 and using $T_{\mbox{\scriptsize eff}}=6070$K, log\,$g$=4.00, and $\xi=2.00$\,\kms.}
}
\end{table*}

Stellar parameters and LTE abundances for Fe, Ti, Sc, and Mg, Ca and Ba for the individual stars can be found in Table\,\ref{Tab:Full-Results}. In the table, the mean abundances for each group are given and are labelled with the subscript \emph{mean}. We have also derived average abundances from the average spectra created by co-adding the spectra within each evolutionary group. Stellar parameters and abundances derived from these co-added spectra can be found in Table\,\ref{Tab:Mean-Results}. Aside from LTE abundances for Fe, Sc and Ti, we performed NLTE corrections for abundances derived from Mg\,{\scriptsize I}, Ca\,{\scriptsize I} and Ba\,{\scriptsize II} lines. Mg\,{\scriptsize I} is modelled following new NLTE calculations by the authors Osorio, Barklem \& Lind (in prep.), Ca\,{\scriptsize I} following \citet*{Mashonkina2007} and Ba\,{\scriptsize II} following \citet*{Mashonkina1999}. \\

\subsection{Error estimations}
\begin{figure}
\begin{center}
\includegraphics[width=1\columnwidth]{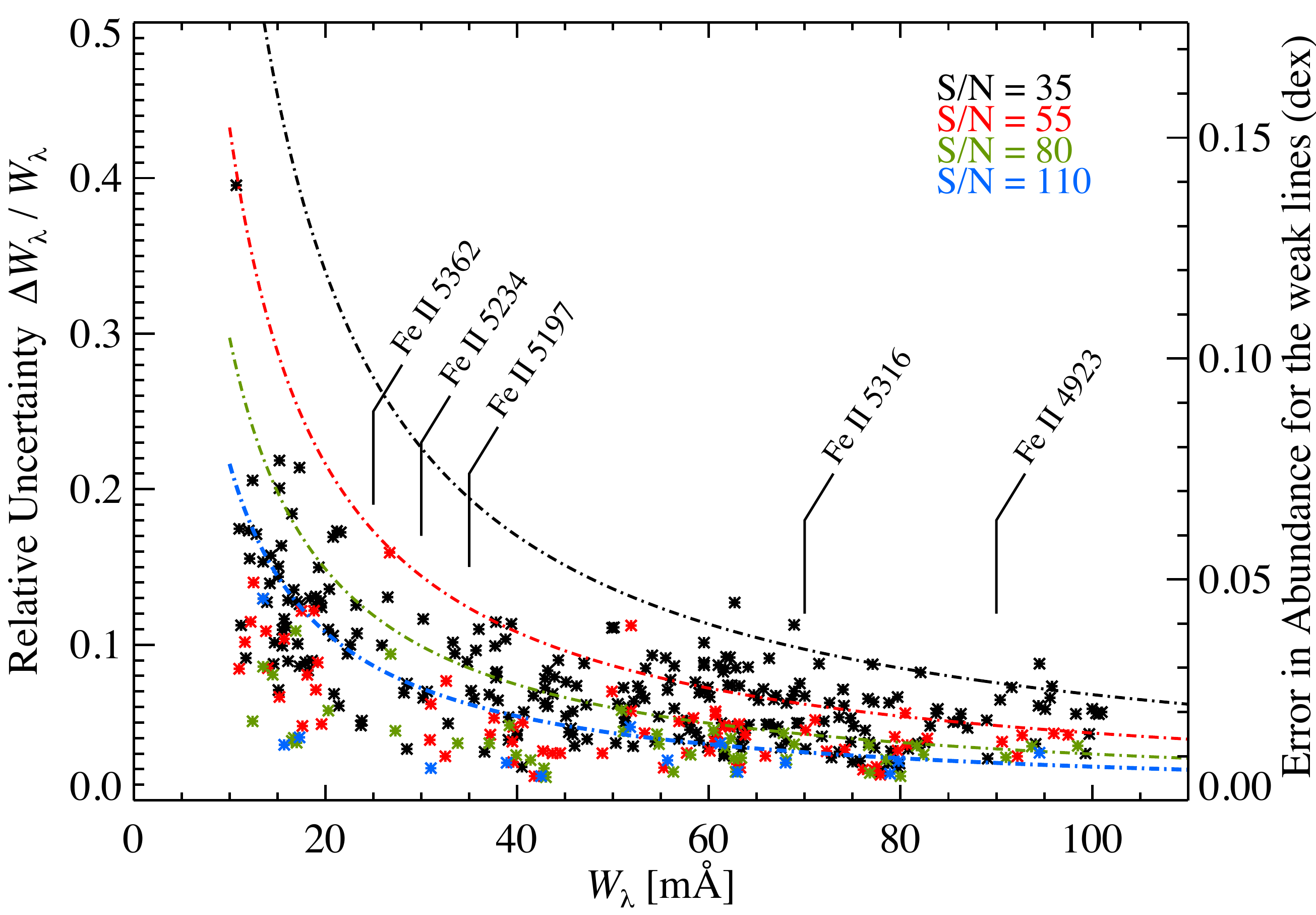}
\caption{Relative uncertainty in EW in function of line strength for different S/N. The line strengths of our five Fe\,{\scriptsize II} lines in the TOP stars are also indicated in the plot. The dot-dashed lines represent the theoretical errors in EW following \citet*{Norris2001} at $\lambda$=5600\AA. On the right axis the corresponding uncertainties in abundance for weak lines are given.}\label{Fig:EWvsER}
\end{center}
\end{figure}

All errors in the abundance tables correspond to line-to-line scatter and do not include the error introduced by uncertainties on stellar parameters. By using weak ionic lines to derive the abundances of Fe, Sc and Ti we can assume errors due to uncertainties in the stellar parameters, to be practically negligible: a shift of 100\,K in $T_{\mbox{\scriptsize eff}}$ translates into 0.04\,dex in log\,$g$ which yields a change in log (abundance) of less than 0.02\,dex. For the other abundances discussed in the paper, the quoted errors do not take into account uncertainties coming from the stellar parameters, (patchy) reddening or (undetected) binarity. \\
For the abundances of individual stars, we quote, in general, the standard deviations around the derived abundances as the error. However, when the abundance is derived from less than three lines, we assume the uncertainty to follow the \citet{Norris2001} formula $\delta\mbox{EW} = \lambda \sqrt{n}/(R[\mbox{S/N}])$ where $n$ is the number of pixels integrated to obtain EW, $R$ the resolving power, and $S/N$ the signal to noise ratio per pixel. When using the  \citet{Norris2001} formula we adopted $R=47000$ and adopted the full-width of the fitted Gaussian measured in pixels for $n$. We investigated the behaviour of the relative uncertainty in EW as given by DAOSPEC as a function of line strength for four different S/N groups. These S/N groups are created by co-adding observations of the same star (NGC\,6397-5281: SGB) in various ways. The observations belong to the dataset described in \citet{Korn2007}. Each observation has a typical S/N of about 35. In total we have 18 observations which can be co-added in different ways to produce a set of spectra that are independent of one another within each S/N sample. As a result, the number of spectra within each S/N sample decreases with increasing S/N. We have 18 spectra with a typical S/N of about 35 (group S/N$_{35}$), six spectra with a typical S/N of 55 (group S/N$_{55}$), three spectra with a typical S/N of 80 (group S/N$_{80}$), and one spectrum with S/N of 110 (group S/N$_{110}$). We then ran DAOSPEC using the same input parameters for all spectra, to place the continuum and measure the EWs of 15 Fe\,{\scriptsize I}. Fig. \ref{Fig:EWvsER} shows the result. In the figure each colour represents the relative uncertainty on the measurement of EW for Fe\,{\scriptsize I} lines by DAOSPEC for a single S/N group. The dash-dotted line represent the theoretical 1-$\sigma$ uncertainty following \citet*{Norris2001} for the different S/N groups. The right axis in the figure gives the translated uncertainty in abundance for weak lines. For convenience we have also marked the typical line strength of the five Fe\,{\scriptsize II} lines in the TOP stars in the cluster NGC\,6752. Given that a typical uncertainty of 10\% in line strength will result in an abundance uncertainty of 0.04\,dex for weak lines, one immediately sees that a S/N higher than 50 is necessary to derive the abundance from weak lines ($\sim30$\,m\AA) with a high enough precision in order to detect abundance differences of the order of 0.1\,dex.\\

Mean abundances in Table\,\ref{Tab:Full-Results} are accompanied by the standard deviation of the mean values. In order to have an analysis as homogeneous as possible, we decided to quote abundances from spectra having roughly the same S/N $\sim 60$. The group mean abundances derived by taking the mean of the abundances of the individual stars within each group are therefore compared to the corresponding average abundances derived from the co-added TOP spectrum. Since the errors associated with the TOP-star abundances correspond to the standard deviation of the individual line abundances, we use the standard error of the mean
\begin{equation*}
\label{Eq:stddev-mean}
\mbox{Std. error of mean} =\frac{ \frac{\sum\sigma}{N}}{\sqrt{N}} = \frac{ \overline{\sigma}}{\sqrt{N}},
\end{equation*}
as error associated with the group mean abundances of the RGB and bRGB, where $\sigma$ is the standard deviation of the mean values and $N$ the number of spectra in each group. The errors associated with the SGB group abundances are again the standard deviation of the individual line abundances since we observed only one SGB star. The final abundances can be found in Table\,\ref{Tab:Final-Results}. 

\begin{table*}
\caption{Mean stellar parameters using ($b-y$) photometry and derived mean abundances for stars in NGC\,6752 from spectra having the same S/N $\sim60$.}\label{Tab:Final-Results}
\centering
\begin{tabular}{lccccccccc}\hline\hline
Group & $T_{\mbox{\scriptsize eff}}$ & log\,$g$ & $\xi$ & [Fe/H] & log\,$\varepsilon$(Mg)\tablefootmark{a} & log\,$\varepsilon$(Sc)\tablefootmark{b}  & log\,$\varepsilon$(Ti)\tablefootmark{c}  & log\,$\varepsilon$(Ca)\tablefootmark{d}  & log\,$\varepsilon$(Ba)\tablefootmark{e} \\ 
& (K) & (cgs) & (km\,s$^{-1}$) & (dex) & NLTE & LTE & LTE & NLTE & NLTE \\ \hline
TOP$_{\mbox{\scriptsize ave}}$ & $6068$ & $4.00$ & $2.00$  & $-1.74\pm0.05$ & $6.01\pm0.07$ & $1.23\pm0.09$ & $3.39\pm0.09$ & $4.88\pm0.07$ & $0.46\pm0.04$ \\
SGB$_{\mbox{\scriptsize mean}}$ & $5793$ & $3.79$ & $1.50$ & $-1.68\pm0.10$ & $6.05\pm0.07$ & $1.21\pm0.09$ & $3.28\pm0.08$ & $4.76\pm0.10$ & $0.44\pm0.03$ \\
bRGB$_{\mbox{\scriptsize mean}}$ & $5256$ & $3.26$ & $1.40$ & $-1.67\pm0.03$ & $6.18\pm0.04$ & $1.32\pm0.05$ & $3.45\pm0.03$ & $4.91\pm0.02$ & $0.48\pm0.01$ \\
RGB$_{\mbox{\scriptsize mean}}$ & $5031$ & $2.58$ & $1.45$ & $-1.66\pm0.03$ & $6.12\pm0.03$ & $1.33\pm0.04$ & $3.45\pm0.02$ & $4.96\pm0.02$ & $0.42\pm0.06$ \\
\hline
$\Delta$[TOP-RGB]  & $1037$ & $1.42$ & & $-0.08\pm0.06$ & $-0.11\pm0.08$ & $-0.10\pm0.10$ & $-0.06\pm0.09$ & $-0.08\pm0.07$ & $0.04\pm0.07$ \\
\hline\hline
\end{tabular}
\tablefoot{
The \emph{average} TOP-star abundances were derived from the co-added TOP-star spectrum while the abundances for the other groups are computed as the \emph{mean} abundance of the abundances derived for the individual stars within each group. Hence, row 1 (TOP$_{\mbox{\scriptsize ave}}$) is identical to row 1 in Table\,\ref{Tab:Mean-Results}. The uncertainties on the abundances for the TOP-group correspond to the standard deviation. For the other groups, they correspond to the standard error of the mean (see text).
\tablefoottext{a}{Based on Mg\,{\scriptsize I} $\lambda\lambda$5528 and 5711. RGB$_{\mbox{\scriptsize ave}}$ Mg abundance derived by disregarding star 548}
\tablefoottext{b}{Based on Sc\,{\scriptsize II} $\lambda\lambda$5031, 5526 and 5657. Note that the Sc abundances for the bRGB and RGB stars can be slightly overestimated due to ignoring HFS.}
\tablefoottext{c}{Based on Ti\,{\scriptsize II} $\lambda\lambda$5129, 5154, 5185, 5226 and 5336.}
\tablefoottext{d}{Based on Ca\,{\scriptsize I} $\lambda\lambda$6169 and 6493.}
\tablefoottext{e}{Based on Ba\,{\scriptsize II} $\lambda$6141. The Ba\,{\scriptsize II} $\lambda$6496 line was disregarded due to telluric contamination. TOP$_{\mbox{\scriptsize ave}}$ Ba abundance derived by disregarding star 3988 and using $T_{\mbox{\scriptsize eff}}=6070$K, log\,$g$=4.00, and $\xi=2.00$\,\kms.}
}
\end{table*}

\subsection{Abundance trends}
Looking at Table\,\ref{Tab:Final-Results} we find systematic weak trends of abundance with evolutionary phase for iron, scandium, magnesium, and calcium. Although the trends are compatible with the null hypothesis it is interesting to note that the derived abundance values for the TOP group are systematic lower than those derived for the RGB group for all investigated species except for barium. 

The iron abundance difference $\Delta$log\,$\varepsilon$(Fe) $=0.08\pm0.06$ is only significant at the 1-$\sigma$ level. The same significance is found in the abundance difference for Ca after correcting for departures from LTE: $\Delta$log\,$\varepsilon$(Ca) $=0.08\pm0.07$.\\

For barium we find a 4$\sigma$ significant trend in LTE: $\Delta$log\,$\varepsilon$(Ba)$_{\mbox{\scriptsize LTE}}=0.28\pm0.07$, but after correcting for departures from LTE the trend disappears to within the errors. Unfortunately, there are no AD predictions for barium. By carefully inspecting the barium lines in the spectra of the individual stars within each group, we found one TOP star (id 3988) which appears to have significantly stronger barium lines compared to the other TOP stars. We therefore disregard this Ba-rich star from the co-addition to create the TOP$_{\mbox{\scriptsize ave}}$ spectrum from which we derived the barium abundance. 

The magnesium abundance difference is, like iron and calcium, only significant at the 1-$\sigma$ level  after applying NLTE corrections: $\Delta$log\,$\varepsilon$(Mg) $=0.11\pm0.08$. However, looking at the difference between bRGB and TOP we find a $2\sigma$ significant trend after correcting for NLTE: $\Delta$log\,$\varepsilon$(Mg) $=0.17\pm0.08$. One has to be careful not to over interpret this result since Mg is affected by internal cluster pollution. We will discuss Mg further below (Sect.\,\ref{sect:Mg}). The abundance differences for titanium and scandium are insignificant but show the same overall behaviour as the other elements discussed. 

\subsection{Identifying distinct stellar populations}\label{sect:Mg}
\citet{Carretta2012} derived Al, Mg and Si abundances for a sample of more than 130 RGB stars in NGC\,6752. They found that the [Al/Fe] abundance ratio in the stars anticorrelates with [Mg/Fe] and [O/Fe] and correlates with [Na/Fe] and [Si/Fe]. Looking at the distribution of stars in the [Al/Fe] abundance ratio plots they found that the stars cluster around three distinct Al values, low, intermediate and high relative to the whole sample. Using Str\"omgren photometry in combination with this finding, these authors were able to identify two stellar generations and three distinct stellar populations in the cluster: a first generation with a primordial population (P) of stars with low Al, Na and high Mg, and a second generation stars which can be split into an intermediate population (I) with intermediate chemical abundances, and an extreme population (E) with high Al, Na and low Mg. Taking advantage of the large sample of derived abundances, they performed a cluster analysis on their data which led to the conclusion that the division is dominated by the Al abundance followed by the Na abundance. There is no strong difference in Mg abundance between first and second generation stars. Only the extreme stellar population seems to have a significantly different Mg abundance from the primordial and intermediate populations. As we are interested in the diffusion trend for Mg, it is important to identify any stars that may belong to the extreme population. The cluster analysis by \citet{Carretta2012} tells us that in order to identify stars belonging to the extreme population, we have to look for stars with significantly lower (higher) Mg (Na) or very significantly higher Al abundances than the mean abundances of the total sample of stars investigated. We therefore deduce Al and Na abundances for our stars where possible. Unfortunately we were unable to deduce Al abundances for the SGB and TOP stars. Sodium NLTE abundances were derived by measuring EWs of the Na\,{\scriptsize I} doublet at 5682-5688\AA\ and comparing them to a grid of NLTE predictions given in \citet{Lind2011}. In the case of the RGB stars, we also included the weak Na\,{\scriptsize I} doublet at 6154-6160\AA. The derived abundances can be found in Table\,\ref{Tab:ExtraAbun}.\\

Based on these abundances we identified one star (id 548) in the RGB group which has a significantly lower Mg and higher Al and Na abundance and can be categorised as an extreme population star. We therefore derived the mean Mg abundance for the RGB group (RGB$_{\mbox{\scriptsize mean}}$) by disregarding the star 548. In the TOP group we found one star (id 4138) with a significantly higher Na abundance than the other stars, however, the Mg abundance is not different from the other stars in the group. We therefore cannot identify it as an extreme star and thus have not excluded it from deriving the mean Mg abundance for the TOP group (TOP$_{\mbox{\scriptsize ave}}$).

\begin{table*}
\caption{Derived elemental abundances for the individual stars in NGC6752.}\label{Tab:ExtraAbun}
\centering
\begin{tabular}{clccccccc}\hline\hline
Group & Star ID & log\,$\varepsilon$(Mg) & log\,$\varepsilon$(Al)  & log\,$\varepsilon$(Na) & log\,$\varepsilon$(Li) & Pop \\ 
 & & NLTE & LTE & NLTE & NLTE & \\ \hline
RGB & 540 & $6.11\pm0.07$ & $5.27\pm0.05$ & $4.74\pm0.10$ & $1.22\pm0.05$ 	& I \\
 	& 548 & $5.92\pm0.09$ & $5.83\pm0.07$ & $4.98\pm0.10$ & $<0.60$\tablefootmark{a} 	& E \\
 	& 566 & $6.12\pm0.07$ & $5.00\pm0.05$ & $4.34\pm0.10$ & $0.95\pm0.08$ 	& P \\
 	& 574 & $6.15\pm0.07$ & $4.83\pm0.10$\tablefootmark{a} & $4.25\pm0.10$ & $0.99\pm0.07$	& P \\
 	& 581 & $6.14\pm0.07$ & $5.35\pm0.05$ & $4.79\pm0.10$ & $0.99\pm0.07$ 	& I \\
 	& 588 & $6.06\pm0.07$ & $5.38\pm0.05$ & $4.74\pm0.10$ & $0.99\pm0.07$ 	& I \\ \hline
bRGB & 1406 & $6.17\pm0.07$ &			    & $4.50\pm0.07$ & $1.01\pm0.13$& P \\
	  & 1483 & $6.11\pm0.07$ & $4.88\pm0.05$ & $4.63\pm0.07$ & $1.13\pm0.10$& \\
	  & 1522 & $6.23\pm0.07$ &			    & $4.44\pm0.07$ & $1.11\pm0.11$& P \\
	  & 1665 & $6.21\pm0.07$ & $5.08\pm0.05$ & $4.71\pm0.07$ & $1.16\pm0.09$ \\ \hline
SGB & 3081 & $6.05\pm0.07$ &  & $4.58\pm0.07$ & $2.11\pm0.05$ \\ \hline
TOP & 3988 & $6.05\pm0.11$ &  & $4.58\pm0.07$ & $2.30\pm0.06$ \\
 	& 4096 & $6.02\pm0.11$ & no reliable & $4.71\pm0.07$ &  $2.34\pm0.07$ \\
 	& 4138 & $6.05\pm0.11$ & abundance & $4.95\pm0.07$ & $2.03\pm0.10$ \\
 	& 4383 & $6.12\pm0.11$ &  & $4.31\pm0.07$ & $2.21\pm0.07$ \\
 	& 4428 & $6.09\pm0.11$ &  & $4.30\pm0.07$ & $2.13\pm0.10$ \\
\hline\hline
\end{tabular}
\tablefoot{
For magnesium two Mg\,{\scriptsize I} lines: 5528 and 5711\AA, for aluminium we used the Al\,{\scriptsize I} doublet at 6696-6698\AA, and for Sodium we used the Na\,{\scriptsize I} doublet at 5682-5688\AA\, for the RGB we also included the Na\,{\scriptsize I} doublet at 6154.2-6160.7\AA. Lithium abundances were derived using the Li\,{\scriptsize I} line at 6707\AA. The last column gives the population: P for primordial, I for intermediate and E for extreme following \citet{Carretta2009b}\\ 
\tablefoottext{a}{Upper limit}
}
\end{table*}

\subsection{Lithium}
The lithium abundance trend with evolutionary phase is heavily affected by dilution of the surface abundance as the stars evolve towards the RGB. As Li is significantly processed in RGB stars and the dilution seems to set in just below the SGB star group \citep{Korn2007}, one has to look at the TOP stars if one wants to deduce original Li abundances for the cluster. We have deduced Li abundances for all our stars using the Li\,{\scriptsize I} line at 6707\AA. The Li abundances are corrected for NLTE following \citet{Lind2009a} and can be found in Table\,\ref{Tab:ExtraAbun}. One immediately notices the large spread in Li abundance of 1.4 dex. The Li abundance drops sharply once the stars evolve onto the RGB as dilution sets in. There is also a clear sign for a Na-Li anticorrelation for two stars in our sample. The RGB star id 548 and TOP star id 4183 both show a significantly higher Na abundance and lower Li abundance than the other stars within the corresponding groups. We therefore disregard these stars in the following discussion.\\
The highest Li abundance, corrected for NLTE, we detect for the TOP stars is log\,$\varepsilon$(Li)$=2.34\pm0.07$ while the lowest RGB Li abundance that we could measure, also corrected for NLTE, is log\,$\varepsilon$(Li)$=0.95\pm0.10$. By comparing our mean RGB Li abundance (disregarding star id 548): log\,$\varepsilon$(Li) $ =1.03\pm0.11$, with the one derived by \citet*{Mucciarelli2012}, who use the temperature scale of \citet*{Gonzalez2009}: log\,$\varepsilon$(Li)$=0.93\pm0.15$, we find good agreement for the RGB mean Li abundances. Even when the spectroscopic temperature scale derived by \citet*{Mucciarelli2012} is used, log\,$\varepsilon$(Li)$=0.88\pm0.15$, the agreement still holds. 

\citet{Pasquini2005} derived LTE Li abundances for nine TOP stars in NGC\,6752 drawn from the study of \citet{Gratton2001} of which we have two stars in common (id 4383 and 4428). Using \citet{Gratton2001} effective temperatures which are roughly 150\,K hotter than ours, they find Li abundances for the TOP stars which are, for 4383 and 4428 respectively, 0.15 and 0.32\,dex higher than what we find. Using  lower $T_{\mbox{\scriptsize eff}}$ values, computed according to the Alonso-scale $(b-y)$ colour and the reddening of $E(b-y) = 0.032$ \citep{Gratton2003}, the Li abundance for star 4383 is in excellent agreement. However, we find a 0.17\,dex discrepancy for star 4428 (discussed further in Sect.\,\ref{Sect:Lithium}).
The mean Li abundance computed from the four TOP stars when disregarding 4138, is log\,$\varepsilon$(Li)$=2.25\pm0.09$ in good agreement with the mean Li abundance log\,$\varepsilon$(Li)$=2.24\pm0.15$ on the hot temperature scale of \citet{Gratton2001}. Given the fairly large star-to-star scatter in lithium abundance, it is also in agreement with the mean Li abundance log\,$\varepsilon$(Li)$=2.15\pm0.14$ derived by \citet{Gratton2001} on the cooler temperature scale of \citet*{Alonso1996}. 

\subsection{Corrections from three-dimensional models}
Recent numerical simulations of convection at the surface of late-type stars have shown that significant structural differences exist between the temperature and density stratifications from three-dimensional (3D), time-dependent, hydrodynamical and classical, one-dimensional (1D), stationary, hydrostatic model stellar atmospheres. In particular, 3D model atmospheres generated with realistic surface convection simulations of metal-poor late-type stars, ranging from dwarfs to giants, are characterised by significantly cooler upper-photospheric temperature stratifications than corresponding 1D model atmospheres constructed for the same stellar parameters \citep{Asplund:1999,Collet:2007}. Such systematic differences between the 3D and 1D temperature stratifications, together with the presence in 3D models of temperature and density inhomogeneities and correlated self-consistent velocity fields, can have substantial effects on the strengths of spectral lines predicted by line formation calculations and, consequently, on elemental abundances derived from spectroscopic analyses.\\

We took into account such effects related to the choice of model stellar atmospheres by performing a differential 3D$-$1D abundance analysis for all elements considered in the present work. More specifically, we recomputed synthetic profiles and EWs for all atomic lines from the various elements using 3D hydrodynamical model atmospheres and assuming an average composition for the stars in NGC6752.  We also synthesized the line profiles using 1D models computed for the same stellar parameters, adjusting the elemental abundances on a line-by-line basis to match the EWs from the 3D line formation calculations. The difference between the 3D and 1D abundances would then give the 3D$-$1D abundance corrections for the individual lines.\\
For the 3D line formation calculations, we used 3D, time-dependent, hydrodynamical \textsc{Stagger} models (Magic, in prep.) with stellar parameters close to the ones of the observed stars in NGC6752, i.e. with effective temperatures between $T_\mathrm{eff}=5000$~K and $6000$~K, surface gravities between $\log{g}=2.5$ and $4.0$ (in cgs units), and metallicities in the range $\mathrm{[Fe/H]}=-1 {\ldots} -2$~dex.\\
For the 1D calculations, we used custom 1D hydrostatic model atmospheres computed with using the same input physics (equation of state, opacities, radiative transfer solver) as in the corresponding \textsc{Stagger} simulations but relying on an implementation of the mixing-length theory for the convective energy transport. In 1D, we also adopted a non-zero value for the micro-turbulence parameter between $\xi=1.7$ and $1.8$~km~s$^{-1}$ for modelling non-thermal Doppler broadening associated with macroscopic flows which are otherwise absent in 1D atmospheres. For both 3D and 1D spectral line formation calculations, we adopted quantum-mechanical calculations by \cite{Barklem:2000} for modelling van der Waals broadening. When these were not available, we relied on the classical description by \cite{Unsold:1955}.\\

From each of the 3D simulations, we selected a subset of 15 time snapshots for the calculations. For each 3D sequence, we computed intensity profiles for the individual spectral lines along 16 inclined directions plus the vertical using \textsc{SCATE} \citep{Hayek:2011}, after which we performed an angular and disk integration and averaged over all selected snapshots to compute the final flux profiles. 
All line profiles were computed under the approximation of LTE.\\
We carried out calculations with a similar setup and line formation code for the 1D model atmospheres, adjusting each time the elemental abundance to match the EWs from the 3D calculations and thus compute the differential 3D$-$1D LTE abundance corrections.
Finally, we linearly interpolated in stellar parameter space the abundance corrections computed for the 3D models from the \textsc{Stagger} grid (and their 1D counterparts) to the stellar parameters representative of the four specific categories of cluster stars considered in this study (TOP, SGB, bRGB, and RGB).\\

The 3D$-$1D LTE abundance corrections for the four groups of stars are listed in Table \ref{tab:abund_corr}. We compare 1D and 3D LTE abundances in Table \ref{tab:1D-3D_abund}. The 3D LTE abundances are derived by applying 3D$-$1D LTE abundance corrections to the measured 1D LTE line abundances and taking the mean. Beside an overall shift to higher abundances (except for Li), the 3D--1D corrections seem to increase the trend slightly for Fe and Ca and significantly for Sc and Ti. Mg seems unaffected beside a shift to higher abundances. From these results we conclude that the abundance trends as such are probably not an artefact of the modelling in 1D.

\begin{table*}
\caption{3D$-$1D LTE abundance corrections for individual lines and for the four groups of cluster stars considered in this work: TOP ($T_\mathrm{eff}$~[K] / $\log{g}$~[cgs] / $[\mathrm{Fe/H}]$~[dex] / $\xi_\mathrm{1D}$~[km~s$^{-1}$] = $6000$/$4.00$/$-1.80$/$2.0$), SGB ($5700$/$3.80$/$-1.70$/$1.5$), bRGB ($5500$/$3.30$/$-1.70$/$1.4$), RGB ($5000$/$2.50$/$-1.70$/$1.45$).}
\label{tab:abund_corr}
\centering
\renewcommand{\footnoterule}{}  
\begin{tabular}{rccccccc}
\hline
\hline
\noalign{\smallskip}
Line & $\chi$ & $\log gf$ &
$\log\varepsilon_\mathrm{3D}$ [dex] &
\multicolumn{4}{c}{$\log\varepsilon_{\mathrm{3D}-\mathrm{1D,LTE}}$ [dex] } \\
 & & & & TOP & SGB & bRGB & RGB \\
\noalign{\smallskip}
\hline
\noalign{\smallskip}
\ion{Li}{I} 6707.8 & 0.000 &  0.174 &  2.000 & -0.230 & -0.241 & -0.385 & -0.270 \\
\ion{Na}{I} 5682.6 & 2.102 & -0.706 &  4.500 &	 ---   & 0.033 & 0.010 & 0.023 \\
\ion{Na}{I} 5688.2 & 2.104 & -0.452 &  4.500 &  0.040 & 0.035 & 0.012 & 0.028 \\
\ion{Na}{I} 6154.2 & 2.102 & -1.547 &  4.500 &  0.030 & 0.026 & 0.001 & 0.012 \\
\ion{Na}{I} 6160.7 & 2.104 & -1.260 &  4.500 &  0.031 & 0.027 & 0.001 & 0.013 \\
\ion{Mg}{I} 5528.4 & 4.346 & -0.498 &  6.000 &  0.121 & 0.080 & 0.033 & 0.122 \\
\ion{Mg}{I} 5711.1 & 4.346 & -1.724 &  6.000 &  0.048 & 0.040 & 0.030 & 0.057 \\
\ion{Al}{I} 6696.0 & 3.143 & -1.347 &  5.000 &	 ---   &	 ---   & 0.008 & 0.021 \\
\ion{Ca}{I} 6169.1 & 2.523 & -0.540 &  4.800 &  0.029 & 0.009 & -0.037 & 0.045 \\
\ion{Ca}{I} 6493.8 & 2.521 & -0.109 &  4.800 &  0.063 & 0.008 & -0.078 & 0.084 \\
\ion{Sc}{II} 5031.0 & 1.357 & -0.400 &  1.200 &  0.062 & 0.034 & 0.098 & 0.158 \\
\ion{Sc}{II} 5526.8 & 1.768 &  0.020 &  1.200 &  0.078 & 0.038 & 0.104 & 0.179 \\
\ion{Sc}{II} 5657.9 & 1.507 & -0.603 &  1.200 &  0.043 & 0.024 & 0.077 & 0.127 \\
\ion{Ti}{II} 5129.1 & 1.892 & -1.240 &  3.400 &  0.107 & 0.057 & 0.165 & 0.245 \\
\ion{Ti}{II} 5154.1 & 1.566 & -1.780 &  3.400 &  0.087 & 0.042 & 0.121 & 0.214 \\
\ion{Ti}{II} 5185.9 & 1.893 & -1.350 &  3.400 &  0.090 & 0.052 & 0.137 & 0.237 \\
\ion{Ti}{II} 5226.5 & 1.566 & -1.230 &  3.400 &  0.171 & 0.087 & 0.160 & 0.313 \\
\ion{Ti}{II} 5336.8 & 1.582 & -1.630 &  3.400 &  0.105 & 0.055 & 0.126 & 0.234 \\
\ion{Ba}{II} 6141.7 & 0.704 & -0.076 &  0.500 &  0.164 & 0.064 & 0.080 & 0.332 \\
\ion{Ba}{II} 6496.9 & 0.604 & -0.377 &  0.500 &  -0.012 & 0.008 & 0.038 & 0.312 \\
\ion{Fe}{II} 4923.9 & 2.891 & -1.530 &  6.000 &  0.514 & 0.282 & 0.307 & 0.370 \\
\ion{Fe}{II} 5197.6 & 3.230 & -2.348 &  6.000 &  0.206 & 0.146 & 0.239 & 0.283 \\
\ion{Fe}{II} 5234.6 & 3.221 & -2.223 &  6.000 &  0.183 & 0.158 & 0.253 & 0.325 \\
\ion{Fe}{II} 5264.8 & 3.230 & -3.250 &  6.000 & 	 ---   & 0.055 & 0.108 & 0.090 \\
\ion{Fe}{II} 5284.1 & 2.891 & -3.200 &  6.000 &	 ---   & 0.064 & 0.139 & 0.156 \\
\ion{Fe}{II} 5316.6 & 3.153 & -1.890 &  6.000 &  0.247 & 0.144 & 0.200 & 0.240 \\
\ion{Fe}{II} 5325.6 & 3.221 & -3.324 &  6.000 &	 ---   & 0.021 & 0.107 & 0.082 \\
\ion{Fe}{II} 5362.9 & 3.199 & -2.570 &  6.000 &  0.163 & 0.103 & 0.188 & 0.274 \\
\ion{Fe}{II} 5425.3 & 3.199 & -3.390 &  6.000 &	 ---   &	 ---   & 0.104 & 0.077 \\

\noalign{\smallskip}
\hline \hline
\end{tabular}
\end{table*}
 
\newpage
\begin{landscape}
\begin{table}
\caption{1D and 3D LTE abundances for the four groups of cluster stars considered in this work.}
\label{tab:1D-3D_abund}
\centering
\renewcommand{\footnoterule}{}  
\begin{tabular}{lcccccccccc}
\hline
\hline
\noalign{\smallskip}
Group  & \multicolumn{2}{c}{\FeH\ [dex]} & \multicolumn{2}{c}{$\log\varepsilon$(Mg) [dex]} & \multicolumn{2}{c}{$\log\varepsilon$(Sc) [dex]}
 & \multicolumn{2}{c}{$\log\varepsilon$(Ti) [dex]} & \multicolumn{2}{c}{$\log\varepsilon$(Ca) [dex]} \\
 & 1D & 3D & 1D & 3D & 1D & 3D & 1D & 3D & 1D & 3D \\
\noalign{\smallskip}
\hline
\noalign{\smallskip}
TOP$_{\mbox{\scriptsize ave}}$ & $-1.74\pm0.05$ & $-1.47\pm0.04$ & $6.08\pm0.04$ & $6.17\pm0.07$ & $1.23\pm0.09$ & $1.29\pm0.09$ & $3.39\pm0.09$ & $3.50\pm0.12$ & $4.93\pm0.07$ & $4.98\pm0.07$ \\
SGB$_{\mbox{\scriptsize mean}}$ & $-1.68\pm0.10$ & $-1.51\pm0.12$ & $6.07\pm0.10$ & $6.13\pm0.10$ & $1.21\pm0.09$ & $1.24\pm0.10$ & $3.28\pm0.08$ & $3.34\pm0.12$ & $4.76\pm0.07$ & $4.77\pm0.07$ \\
bRGB$_{\mbox{\scriptsize mean}}$ & $-1.67\pm0.03$ & $-1.43\pm0.06$ & $6.25\pm0.04$ & $6.28\pm0.05$ & $1.32\pm0.05$ & $1.41\pm0.05$ & $3.45\pm0.03$ & $3.59\pm0.05$ & $4.93\pm0.06$ & $4.87\pm0.06$ \\
RGB$_{\mbox{\scriptsize mean}}$ & $-1.66\pm0.04$ & $-1.36\pm0.04$ & $6.22\pm0.04$ & $6.31\pm0.02$ & $1.33\pm0.04$ & $1.48\pm0.04$ & $3.45\pm0.02$ & $3.70\pm0.02$ & $4.97\pm0.04$ & $5.03\pm0.04$ \\
\hline
$\Delta$[TOP-RGB]$_{\mathrm{3D}}$ & & $-0.11\pm0.06$ & & $-0.14\pm0.07$ & & $-0.19\pm0.10$ & & $-0.20\pm0.12$ & & $-0.05\pm0.08$ \\
\hline\hline
\end{tabular}
\tablefoot{
The \emph{average} 1D TOP-star abundances were derived from the co-added TOP-star spectrum while the abundances for the other groups are computed as the \emph{mean} abundance of the abundances derived for the individual stars within each group. The 3D abundances are derived by applying the corrections given in Table\,\ref{tab:abund_corr}. The uncertainties on the abundances for the TOP-group correspond to the standard deviation. For the other groups, they correspond to the standard error of the mean (see text).}
\end{table}
\end{landscape} 

\section{Discussion}
\subsection{Comparison with diffusion models}
\begin{figure*}[ht]
\centering
\includegraphics[width=\columnwidth]{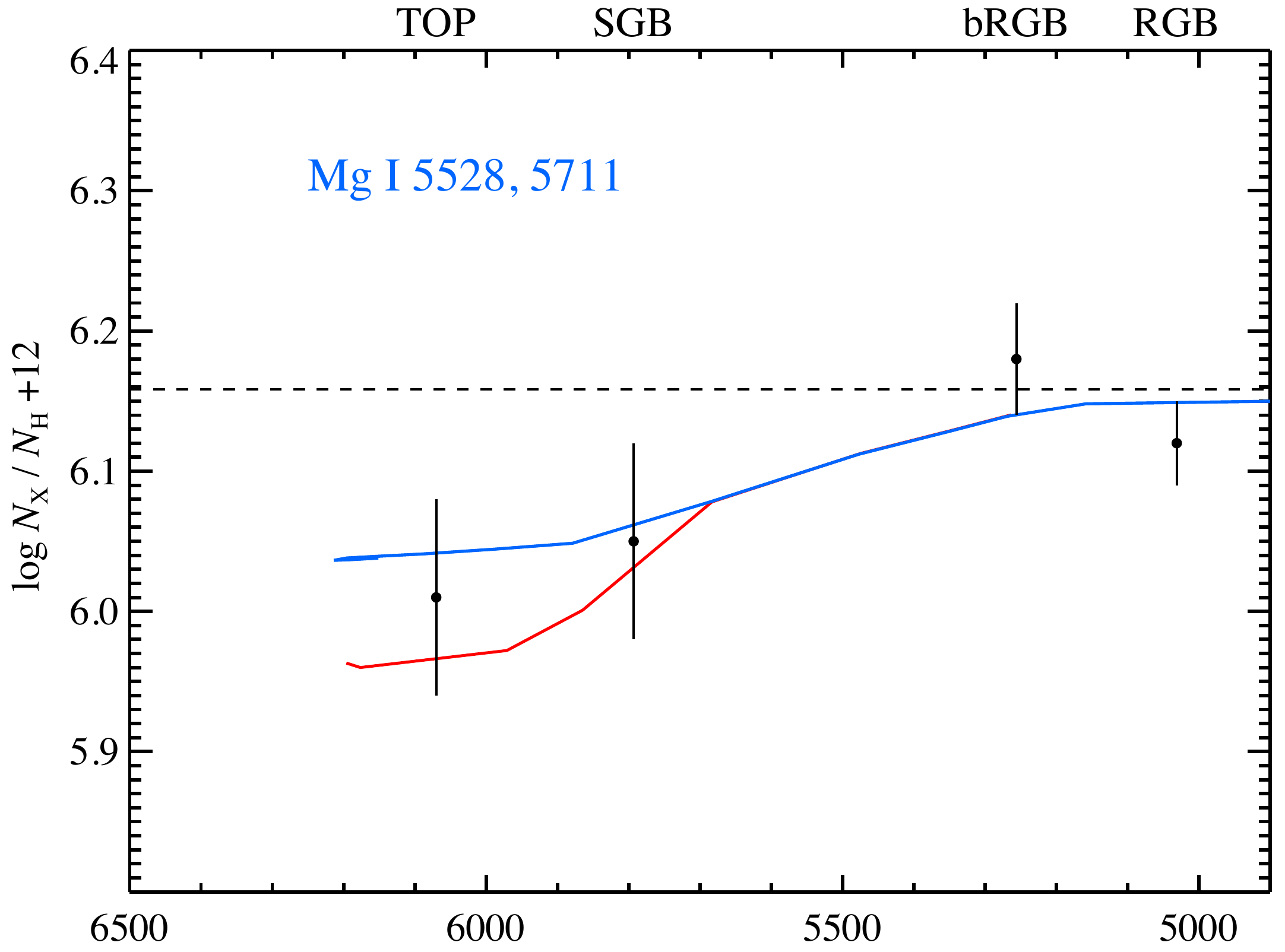}
\includegraphics[width=0.95\columnwidth]{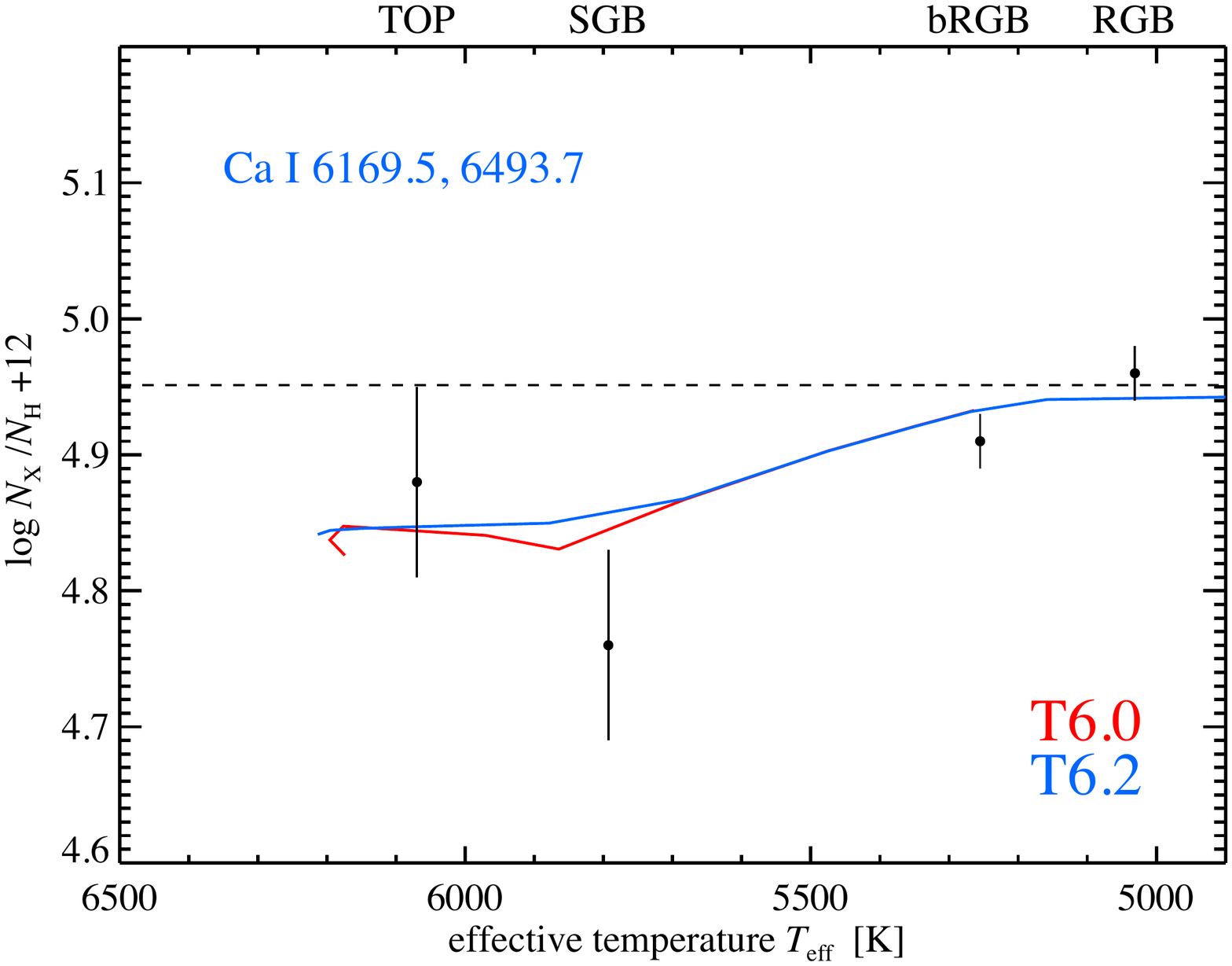}\\
\includegraphics[width=\columnwidth]{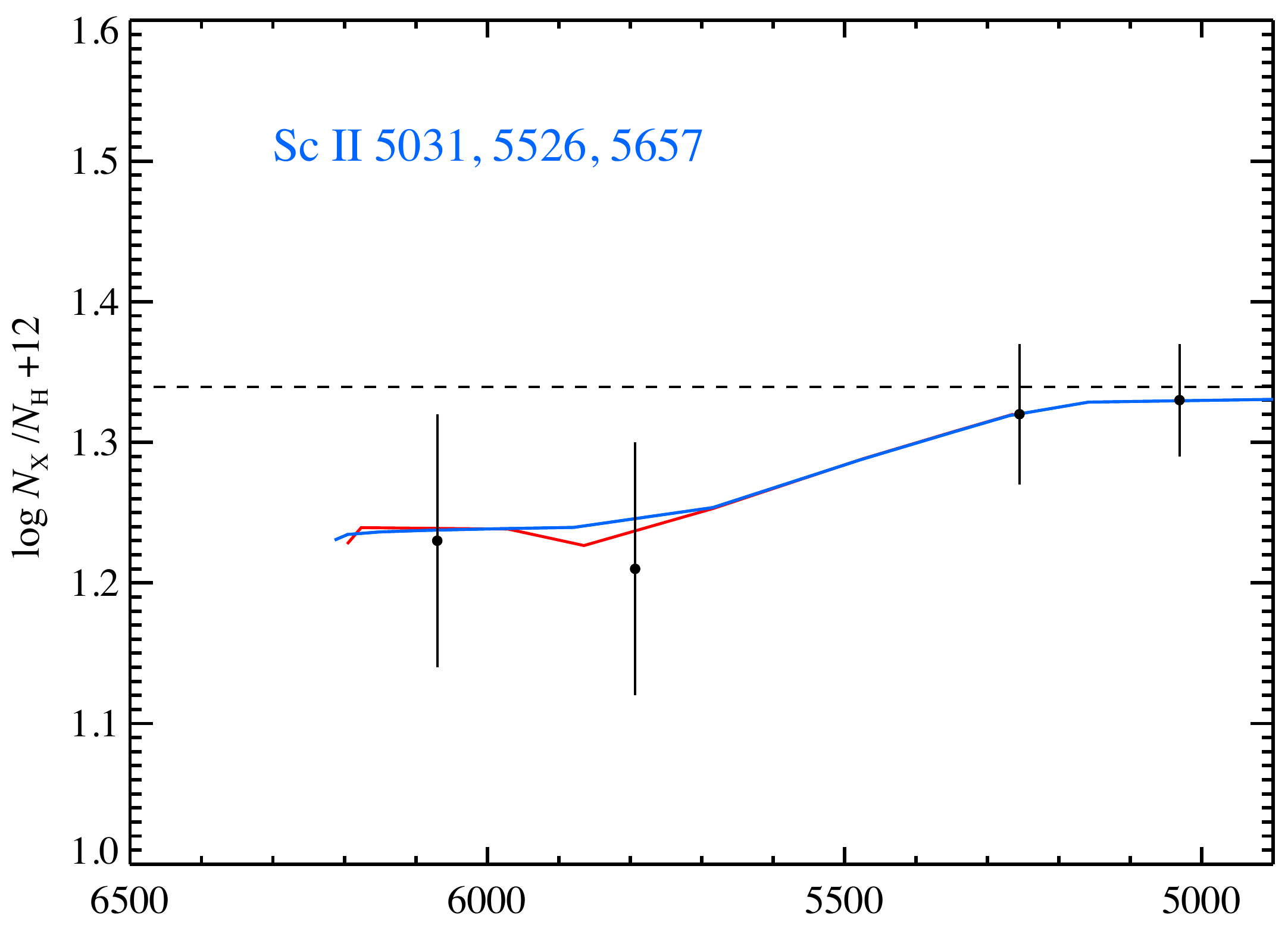}
\includegraphics[width=0.95\columnwidth]{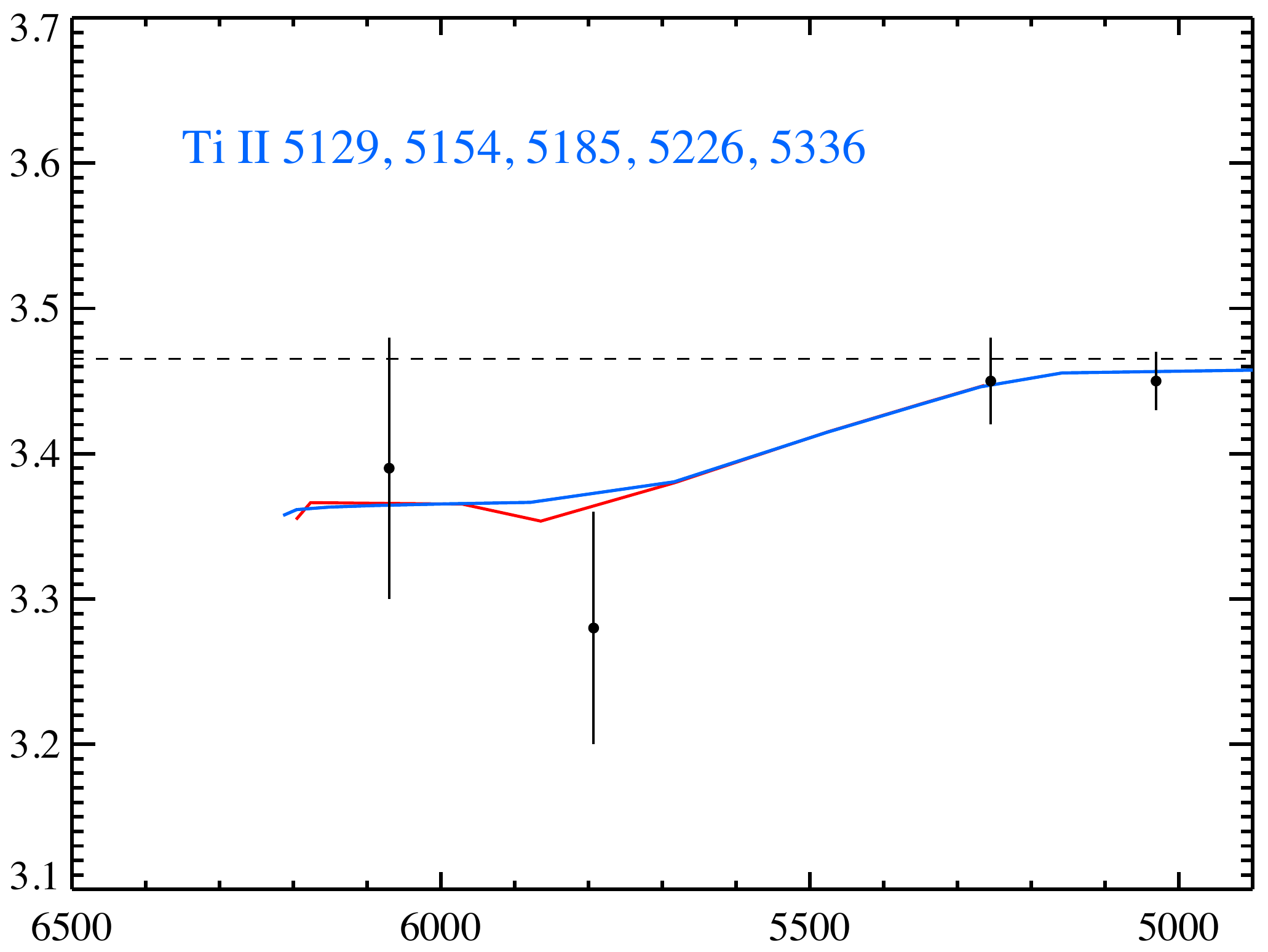}\\
\includegraphics[width=\columnwidth]{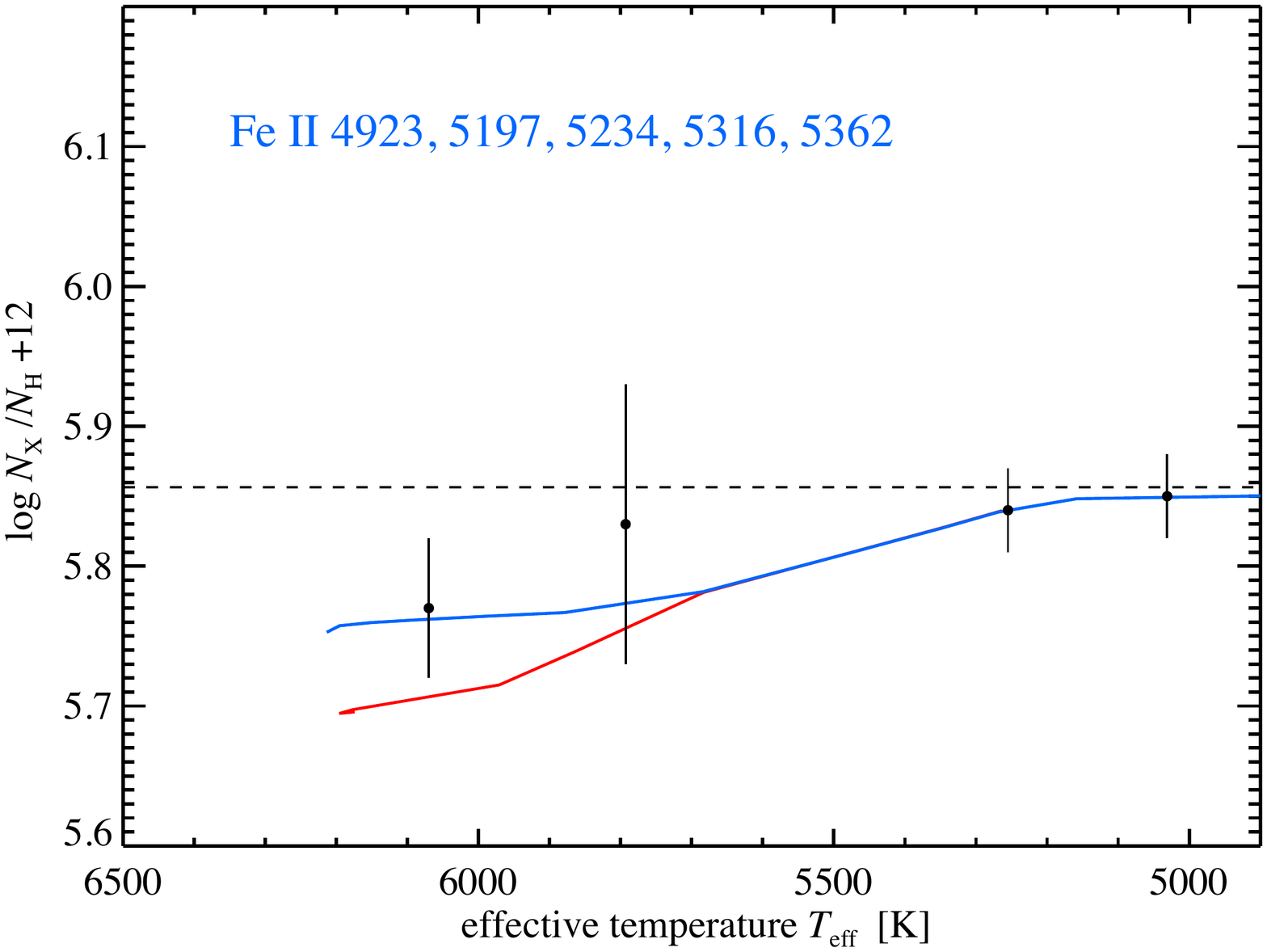}
\includegraphics[width=0.97\columnwidth]{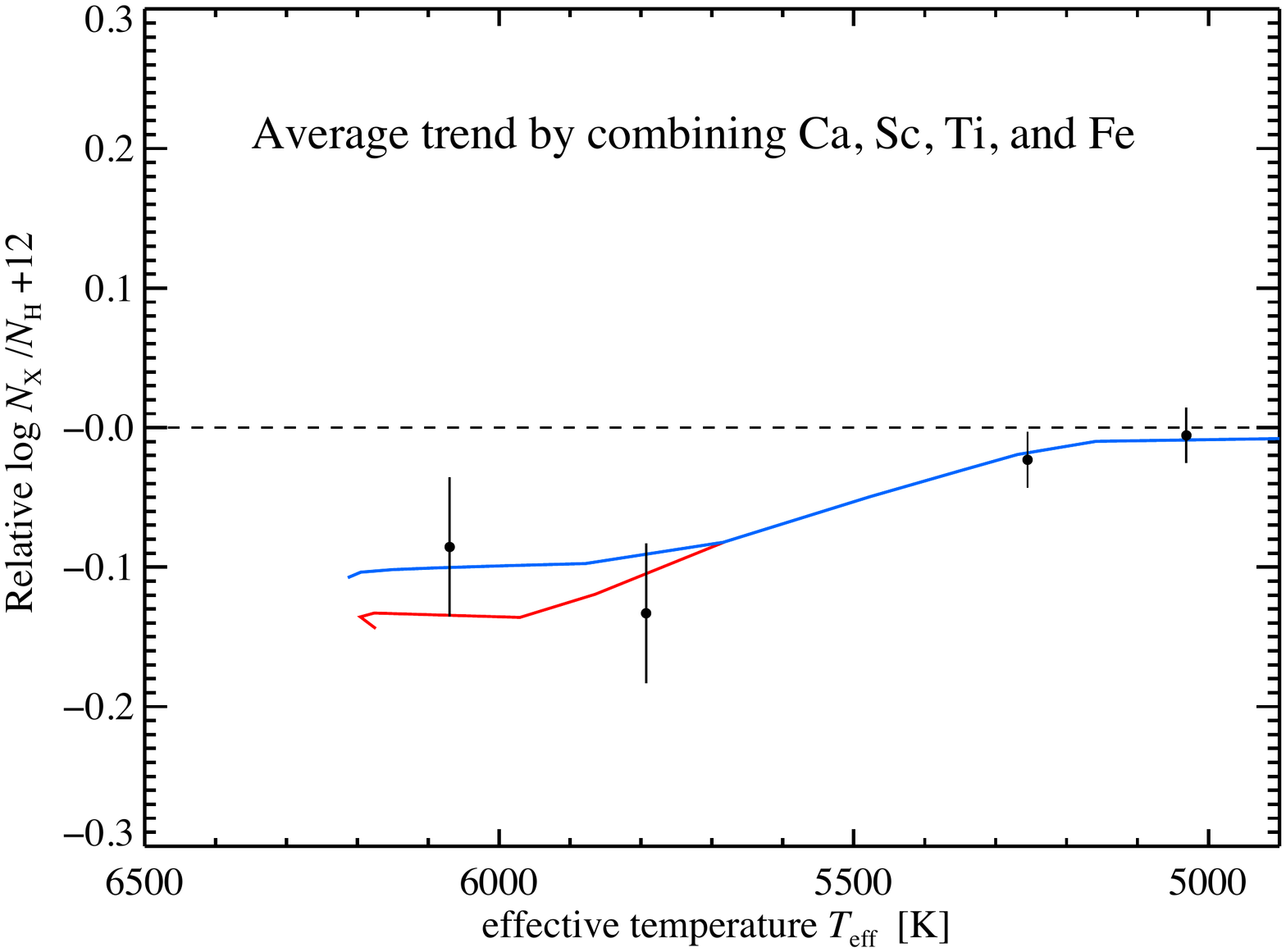}
\caption{Observed abundance trends with evolutionary stage. Each point represents the abundance derived from the co-added spectra within each group as a function of $T_{\mbox{\scriptsize eff}}$. The curves through the points are predictions from stellar evolution models including AD and AddMix with two different efficiencies, at 13.5\,Gyr. The blue curve represents high-efficiency mixing ($T_0$=6.20) while the red curve corresponds to low-efficiency mixing ($T_0$=6.00). The \emph{bottom right} panel shows the mean trend by normalising the observations and models for Ca, Sc, Ti and Fe to the primordial abundances given by the models and afterwards averaging these normalised observations and model trends.
}\label{Fig:trends}
\end{figure*}

We can now compare our abundance variations with stellar-structure models that allow for radial diffusion of chemical elements. For this comparison we use the predictions from AD models of Population II stars by \citet{Richard2002,Richard2005}, which account for all the physics of particle transport that can be modelled from first principles. The models for NGC\,6752 were computed as described in \citet*{Richard2005}. For the models, they use a metallicity of \FeH =--1.6 with an $\alpha$-enhancement of 0.3\,dex compared to the solar mixture and isochrones which were computed at 13.5\,Gyr. Transport of unknown physical origin below the outer convection zone (AddMix) is the main free parameter in these models. This transport is parameterised by a (decreasing) function of temperature and density. To reproduce the observation of the ``Spite plateau" of lithium \citep{Spite1982} with models including AD, one needs to invoke AddMix to counteract the too strong downward depletion of lithium at the surface of the hottest stars. In Fig.\,\ref{Fig:trends} predictions from two models are overplotted on top of our observations. The models implement different values of the reference temperature $T_0$ (log $T$ = 6.00/6.20), which controls the efficiency of the mixing, and use an AddMix turbulent diffusion coefficient $D_{T}$ which is chosen to be 400 times the atomic-diffusion coefficient for helium at the reference temperature $T_0$ and varying as $\rho^{-3}$ \citep*{Michaud1991}. The red curves in Fig.\,\ref{Fig:trends} represent AD models with AddMix at low efficiency, denoted by T6.00, while the blue curves characterise models with AddMix at high efficiency, denoted by T6.20. We slightly adjusted the absolute abundances predicted by the two models, individually for each element. This to establish agreement with our derived abundances in the cooler half of the effective temperature range ($\leqslant$5600\,K), where the two models coincide. As the convective envelope fully encompasses the regions mixed by turbulence, the mixing efficiency is irrelevant for the surface abundances in these evolved stars. The black dashed lines mark the initial abundances of the models. \\
Higher efficiency mixing results in shallower trends for most elements, including iron and magnesium, when comparing TOP and RGB abundances. The explanation for this behaviour has to do with the fact that radiative acceleration is weaker than gravity at the bottom of the zone homogenised by mixing. Increasing the efficiency of AddMix reduces the effect of gravitational settling. At $T_{\mbox{\scriptsize eff}} \leqslant 5100$\,K the convective zone is sufficiently deep to erase the effects of AD.  For calcium, scandium and titanium, however, the situation is different. Radiative acceleration on these elements is larger than gravity at the bottom of the zone homogenised by mixing. This then results in slightly higher abundances at the turnoff for the model with AddMix at low efficiency. This is because increasing AddMix efficiency extends the zone homogenised by mixing to deeper layers in the star where the radiative acceleration on elements, such as Ca, Sc, and Ti, becomes weaker than gravity, making the trends mildly steeper. Also, there might be a possible influence of mass loss since supported elements are expelled through the surface, while sinking elements encounter advection by the wind. Ca, Sc, and Ti become more overabundant for a short period of time because the wind has an effect just below the Surface Convection Zone, where radiative acceleration equals gravitational settling (see Fig.\,4 in \citealt{Vick2013}); therefore, instead of accumulating below the surface-convection zone, they are pushed into it. \citet{Vick2013} investigated the influence of mass loss on the concentrations of metals in the interior of stars. They compared stellar-structure models with mass loss with models including AD with AddMix and found that larger differences appear in the interior concentrations of metals (see their Fig.\,9). Comparing their results with observations by \citet{Nordlander2012} of stars in the GC NGC\,6397 they find that models with mass loss agree slightly better with subgiant observations than those with turbulence. Lower RGB stars instead favor the models with turbulence. \\

The T6.00 model was adopted by \citet{Korn2007} and \citet{Nordlander2012} as the best fit to the observations for the cluster NGC\,6397 at a metallicity \FeH$\sim$\,--2.1, while the T6.20 model seems to best match the flatter trends found for the cluster NGC\,6572 at metallicity \FeH\,=\,--1.6. In principle one could question the significance of the shallow trends but the fact remains that we find lower abundances for the TOP stars compared to the RGB stars for all investigated elements except barium. 
To add to the significance of the trends, we compute a mean trend by normalising the observations and models to the original abundances given by the models and afterwards averaging these normalised observations and model trends. The result can be seen in the bottom right panel of Fig.\,\ref{Fig:trends}. We have also calculated the weighted mean of these points and calculated the $\chi^2$ with respect to this value. We find that there is only a 7\% probability that the points are compatible with a null-trend centred on the weighted mean. The combined abundance trend is thus significant at the $\sim$ 2-$\sigma$ level and likely not the result of random scatter around a mean value due to measurement errors.

\subsection{Consistency check for the Fe abundance difference}
As an alternative method we derived stellar Fe abundances using a tool based on the software package {\textrm Spectroscopy Made Easy} (SME) by \citet{Valenti1996}. The package may be used to determine stellar parameters by matching observed spectra with synthetic spectra generated from a model-atmosphere grid. As input, SME requires a line list, line and continuum masks, and the observed spectrum with corresponding S/N ratio. SME then determines stellar parameters such as effective temperature, surface gravity, metallicity, micro- and macroturbulence, and individual abundances by matching spectra. \\

For the {\sl Gaia}-ESO public spectroscopic Survey (GES; \citealt{Gilmore2012}), tools for deriving stellar abundances in an automatic way are being developed by different teams involved in the science analysis. One of the tools is based on SME. We used this tool with input line list, line and continuum masks, and setup optimised for GES by M. Bergemann (priv. comm.). For direct comparison with the main analysis of this study, we used the DAOSPEC-normalised spectra as input spectra and chose not to let SME continuum-normalise the spectra again using the continuum masks. Using the stellar parameters found in Table\,\ref{Tab:Full-Results}, we derive the iron abundance from 48 Fe\,{\scriptsize I} and 10 Fe\,{\scriptsize II} lines and we find an iron abundance difference between TOP and RGB stars of $-0.09\pm0.02$.\\

Thus, using two independent methods we find an Fe abundance difference between TOP and RGB stars in NGC\,6752 which is best explained by stellar evolution models including AD and AddMix parametrised by log $T$ = 6.20. It seems that the efficiency of AddMix is stronger in this GC than in the more metal-poor GC NGC\,6397. This is an interesting result as the stars in the two clusters differ only by 0.5\,dex in metallicity.  Across this metallicity regime, AD predictions do not differ significantly. The difference in the abundance trends must thus stem from physical effects not captured by the stellar-structure models employed by us where AddMix is a free parameter.

\subsection{The primordial lithium abundance}\label{Sect:Lithium}
\begin{figure}[ht]
\centering
\includegraphics[width=\columnwidth]{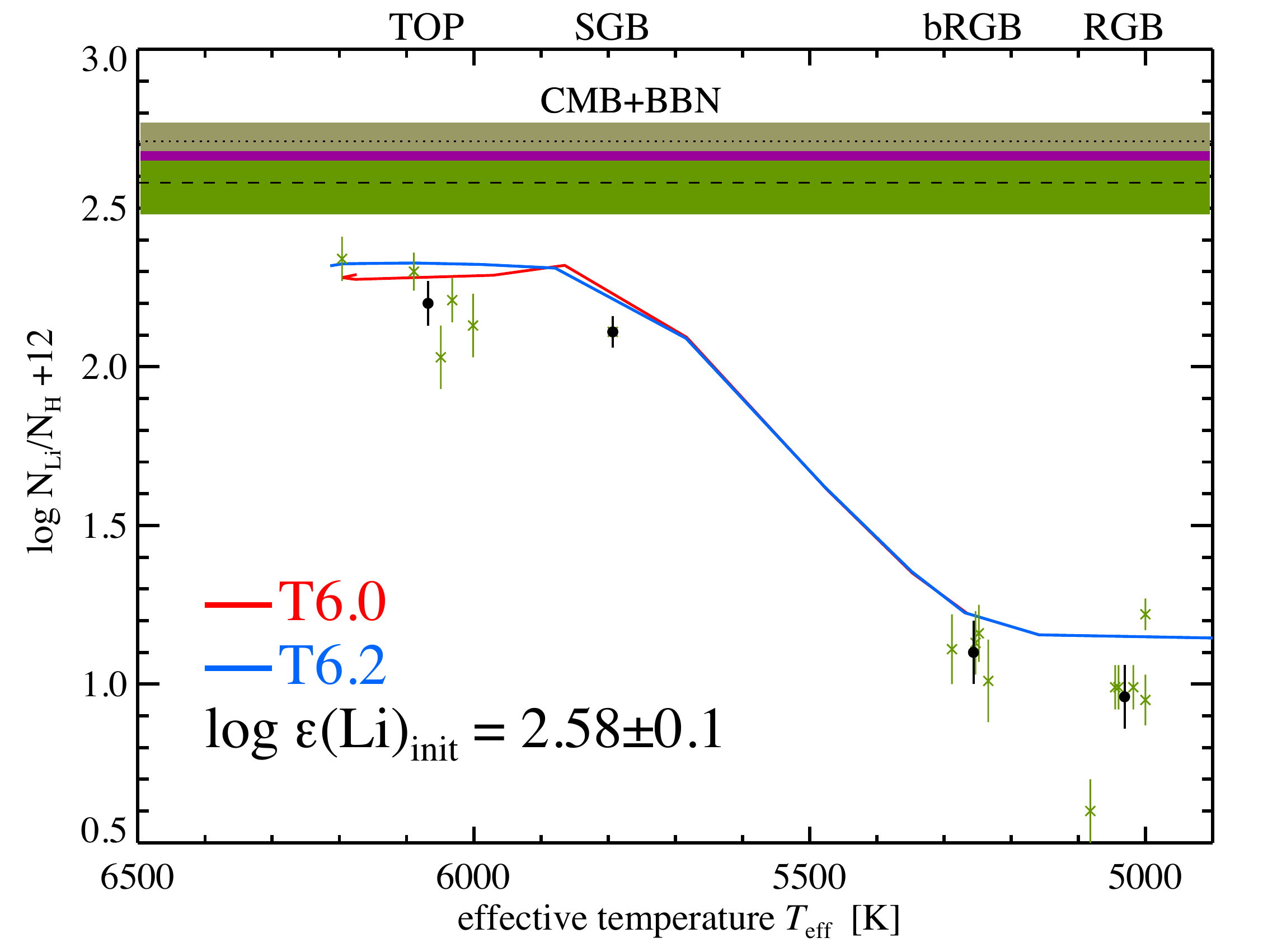}
\caption{\footnotesize Observed abundance trends with evolutionary stage of lithium. Green crosses indicate individual stellar abundances, while bullets represents the abundance derived from the co-added spectra within each group as a function of $T_{\mbox{\scriptsize eff}}$. Solid lines represent predictions from stellar evolution models including AD and AddMix with two different efficiencies, at 13.5\,Gyr. The blue curve represents high-efficiency mixing ($T_0$=6.20) while the red curve corresponds to low-efficiency mixing ($T_0$=6.00). The dashed line gives the original Li abundance for the mean of our two highest detected Li abundances following the AD model with AddMix parameterised by $T_0$=6.20, $\log \varepsilon$(Li)$=2.58\pm0.1$. The dotted line with the olive-brown shaded area shows the predicted primordial lithium abundance [``CMD+BBN": $\log \varepsilon$(Li)$=2.71\pm0.06$; \citealt{Cyburt2010}]. The purple shaded area represent the overlapping 1-$\sigma$ uncertainties between the two values. 
}\label{Fig:Li-diff}
\end{figure}

Our sample of TOP stars seems to include two stars that have significant higher Li abundances than the others (id 3988 and 4096). Using the mean abundance derived from these two stars and correcting for departures from LTE following \citet{Lind2009a}, we can apply the AD correction for lithium. Based on AD models with AddMix characterised by T$6.20$, we find an initial Li abundance of log $\varepsilon$(Li) = $2.58\pm0.10$ dex. This value lies within the mutual 1-$\sigma$ errors with the Wilkinson Microwave Anisotropy Probe (WMAP)-calibrated Big Bang Nucleosynthesis (BBN) predictions by \citet{Cyburt2010}: log\,$\varepsilon$(Li)$_{BBN}$ = $2.71\pm0.06$. This is shown in Fig.\,\ref{Fig:Li-diff}.\\

Our analysis of star 4428 (with a Li abundance of $2.50\pm0.04$ according to \citet{Pasquini2005}) indicates a lower-than-average Li abundance of $2.18\pm0.10$ in LTE. The difference with respect to \citet{Pasquini2005} can be fully explained by the difference in $T_{\mbox{\scriptsize eff}}$ (--225\,K) and in measured line strength (--9\,m\AA). From this work, it seems unlikely that stars with Li as high as 2.50 in log\,$\varepsilon$(Li) exist in NGC 6752. Rather, this cluster displays an upper envelope in Li compatible with field stars of the same metallicity \citep{Nissen2012}.

\section{Conclusions}
We have identified abundance trends for the elements Mg, Ca, Sc, Ti, Fe, and Ba between groups of stars from the TOP to the RGB in the globular cluster NGC\,6752. The six elements represent groups of elements which are affected differently by AD and AddMix (due to the element-specific interplay of gravitational settling and radiative acceleration). The trends can be explained with the stellar models of \citet{Richard2005} which include atomic diffusion and additional mixing. The optimal model indicates an initial lithium abundance of $\log \varepsilon$(Li)$_{\mbox{\scriptsize init}} = 2.58\pm0.10$. This is compatible with the predicted primordial abundance based on WMAP-7 data (\citet{Cyburt2010}, $\log \varepsilon$(Li)$_{\mbox{\scriptsize CMB+BBN}} = 2.71\pm0.06$).\\

Although the significance of the trends is weak, they seem to indicate that AD is operational along the evolutionary sequence of NGC\,6752. However, to explain the observed trends one needs to include more efficient AddMix than in NGC\,6397 as studied by \citet{Korn2007} and \citet{Nordlander2012}. This is interesting, as the two clusters differ by only 0.5\,dex in metallicity. Judging from these results, it seems that there is an interplay between metallicity and the efficiency of AddMix (evidence for this is weaker when the 3D--1D abundance corrections are used). If these results reflect trends in nature, then the expectation is that the diffusion signature will eventually disappear in stars with higher metallicities as AddMix becomes more efficient and suppresses the abundance differences between groups of stars in different evolutionary phases. This picture may then explain why \citet{Mucciarelli2011} need an AD model with very efficient mixing (T6.30) to make the lithium abundances compatible with WMAP-calibrated BBN in the GC M4 (\FeH $\sim$ --1.1). The fact that they did not observe a trend in iron in M4 while the T6.30 model by O. Richard predicts a small (0.1\,dex) trend, could be due to simplifying assumptions (1D, LTE) in their analysis of the stellar spectra. Studying more metal-poor clusters, one would expect to find stronger diffusion trends as the effect of AddMix decreases. GCs such as M92 and M30 are prime targets in this respect.\\ 


\begin{acknowledgements}
PG, AK and PB thank the European Science Foundation for support in the framework of EuroGENESIS. AK acknowledges support by the Swedish National Space Board.
OR acknowledge HPC@LR and Calcul Qu\'eb\'ec for providing the computational resources required for stellar evolutionary computations. OR also acknowledges the financial support of Programme National de Physique Stellaire (PNPS) of CNRS/INSU.
FG acknowledges The Danish National Research Foundation for providing funding for the Stellar Astrophysics Centre. The research is supported by the ASTERISK project (ASTERoseismic Investigations with SONG and Kepler) funded by the European Research Council (Grant agreement no.: 267864).
LM thanks the Swiss National Science Foundation (SCOPES project No.~IZ73Z0-128180/1) for partial support of this study. 
P.S.B is a Royal Swedish Academy of Sciences Research Fellow supported by a grant from the Knut and Alice Wallenberg Foundation.
YO and PB gratefully acknowledge the support of G\"oran Gustafssons Stiftelse.

\end{acknowledgements}

\bibliographystyle{aa}
\bibliography{allreferences}

\end{document}